\documentclass[%
reprint,
 amsmath,amssymb,
 aps,
]{revtex4-1}

\usepackage{graphicx}
\usepackage{dcolumn}
\usepackage{bm}

\usepackage{units}
\usepackage{hyperref}

\begin{document}


\title{Hydrodynamic, Optically-Field-Ionized (HOFI) Plasma Channels}
\author{R. J. Shalloo$^1$\thanks{R.J.Shalloo was responsible for experiment, C. Arran for theory and simulations}}
\author{C. Arran$^1$}
\author{L. Corner$^1$}
\author{J. Holloway$^1$}
\author{J. Jonnerby$^1$}
\author{R. Walczak$^1$}
\author{H. M. Milchberg$^2$}
\author{S. M. Hooker$^1$}
\email{simon.hooker@physics.ox.ac.uk}

\affiliation{$^1$John Adams Institute for Accelerator Science and Department of Physics, University of Oxford, Denys Wilkinson Building, Keble Road, Oxford OX1 3RH, United Kingdom}
\affiliation{$^2$Institute for Research in Electronics and Applied Physics, University of Maryland, College Park, Maryland 20742, USA}

\date{\today}

\begin{abstract}
We present experiments and numerical simulations which demonstrate that fully-ionized, low-density plasma channels can be formed by hydrodynamic expansion of plasma columns produced by optical field ionization (OFI). The laser energy required to generate these channels is modest: of order \unit[1]{mJ} per centimetre of channel. Simulations of the hydrodynamic expansion of plasma columns formed in hydrogen show the generation of \unit[200]{mm} long plasma channels with axial densities of order $n_\mathrm{e}(0) = \unit[1 \times 10^{17}]{cm^{-3}}$ and lowest-order modes of spot size $W_\mathrm{M} \approx \unit[40]{\mu m}$. The simulations are confirmed by experiments with a conventional lens which show the formation of short plasma channels with  $\unit[1.5 \times 10^{17}]{cm^{-3}} \lesssim n_\mathrm{e}(0) \lesssim \unit[1 \times 10^{18}]{cm^{-3}}$ and  $\unit[61]{\mu m} \gtrsim W_\mathrm{M} \gtrsim  \unit[33]{\mu m}$. Low-density plasma channels of this type would appear to be well-suited as multi-GeV laser-plasma accelerator stages capable of long-term operation at high pulse repetition rates.

This article was published in \text{Physical Review E} \textbf{97}, 053203 on 7 May 2018. \href{https://doi.org/10.1103/PhysRevE.97.053203}{DOI: 10.1103/PhysRevE.97.053203}

\noindent \copyright 2018 American Physical Society.
\end{abstract}

\pacs{Valid PACS appear here}

\maketitle

\section{Introduction}
The interaction of intense laser pulses with plasma results in a wealth of phenomena, with many important applications, such as the generation of coherent X-rays or acceleration of charged particles. In many of these cases it would be desirable to increase the laser-plasma interaction distance beyond the limits set by diffraction or refraction of the laser beam. The development of methods for guiding intense laser pulses through plasma is therefore important for many areas of plasma physics.

Laser-driven plasma accelerators are of particular relevance to the present work. In these, an intense laser pulse drives a trailing density wave, within which are formed very large longitudinal and transverse electric fields. Accelerating fields of order $\unit[100]{GV m^{-1}}$ can be generated \cite{Hooker:2013jk}, which is three orders of magnitude higher than possible with radio-frequency technology. Laser-driven plasma accelerators have generated electron beams with particle energies in the GeV range in accelerator stages only a few centimetres long \cite{Leemans:2006, Kneip:2009, Wang:2013el, Leemans:2014kp}, with bunch durations in the femtosecond range \cite{Buck:2011dg, Lundh:2011b, Heigoldt:2015cd}, and with properties ideal for generating femtosecond duration visible to X-ray pulses  \cite{Schlenvoigt:2008, Fuchs:2009, Kneip:2010,Cipiccia:2011, Phuoc:2012vb, Powers:2013bx, Khrennikov:2015gx}.

Waveguides are essential for laser-plasma accelerators operating in the quasi-linear regime, in which the normalized vector potential $a_0 = e A_0/m_\mathrm{e} c \approx 1$, where $A_0$ is the peak vector potential of the laser field. In this regime the particle energy gain of the accelerator varies \cite{Esarey:2009} as $\Delta W \propto 1 / n_\mathrm{e} \propto \lambda_\mathrm{p}^2$, where $n_\mathrm{e}$ is the electron density, whereas the distance required to reach this energy varies as $L_\mathrm{acc} \approx (1/2) \lambda_\mathrm{p}^3/ \lambda^2 \propto 1 / n_\mathrm{e}^{3/2}$. Hence, compared to a \unit[1]{GeV} stage, a $\unit[10]{GeV}$ accelerator requires a decrease in the plasma density by an order of magnitude, and an increase in the accelerator length by a factor of 30. For example, recent design studies for \unit[5]{GeV} and \unit[10]{GeV} accelerator stages propose $n_\mathrm{e} =\unit[1.8 \times 10^{17}]{cm^{-3}}$, $L_\mathrm{acc} = \unit[118]{mm}$ and $n_\mathrm{e} =\unit[0.96 \times 10^{17}]{cm^{-3}}$, $L_\mathrm{acc} = \unit[600]{mm}$ respectively \cite{Cros:2017, Leemans:2011um}.

To maintain $a_0 \approx 1$ whilst keeping the laser power below the critical power for relativistic self-focusing requires \cite{Leemans:2011um} that the laser spot size $w_0  \lesssim \lambda_\mathrm{p}$. This condition means that the ratio of $L_\mathrm{acc}$ to the Rayleigh range $Z_\mathrm{R} = \pi w_0^2 /\lambda$ increases as the energy gain of the stage is increased, since $L_\mathrm{acc} / Z_\mathrm{R} > \lambda_\mathrm{p} /2  \pi \lambda \propto \sqrt{\Delta W}$.

There is therefore considerable interest in developing waveguides capable of guiding intense laser pulses over distances above \unit[100]{mm}, through plasma with a density of order $\unit[10^{17}]{cm^{-3}}$, and with a matched spot size $W_\mathrm{M} < \lambda_\mathrm{p} \approx \unit[100]{\mu m}$. With potential applications of laser-plasma in mind, it would be highly desirable if the waveguide could operate uninterrupted for extended periods at kilohertz repetition rates.

To date laser-plasma accelerators have employed step-index guiding in hollow capillaries and gradient refractive index guiding in plasma channels. Grazing-incidence guiding in hollow capillary waveguides \cite{Cros:2002} has been shown to guide \cite{Dorchies:1999vb} joule-level pulses with peak intensities above $\unit[10^{16}]{W cm^{-2}}$ over lengths of \unit[100]{mm}, and to generate electron beams in the \unit[100]{MeV} range with improved stability \cite{Hansson:2014fk}. With this approach, laser damage of the capillary, particularly the entrance face, can be a problem if the transverse profile and pointing of the drive laser are not tightly controlled.

An alternative approach is to employ gradient refractive index guiding in a plasma channel, i.e.\ a cylinder of plasma in which the electron density increases --- and hence the refractive index decreases --- with radial distance from the axis. Plasma channels have been produced by: slow electrical discharges in evacuated plastic capillaries \cite{Ehrlich:1996}; fast capillary discharges \cite{Hosokai:2000}; open-geometry discharges \cite{Lopes:2003}; hydrodynamic expansion of laser-heated plasma columns \cite{Durfee:1993, Durfee:1994wz, Durfee:1995gr, Kumarappan:2005du}; and gas-filled capillary discharges \cite{Spence:2000fr, Butler:2002zza}. However, only the last two methods have been used to accelerate electrons \cite{Geddes:2004, Leemans:2006, Karsch:2007, Ibbotson:2010, Rowlands-Rees:2008, Leemans:2014kp, Goers:2014jg}. Plasma acceleration has been demonstrated in gas-filled capillary discharge waveguides up to \unit[90]{mm} long  \cite{Leemans:2014kp}; it has also been shown that gas-filled capillary discharges can operate at repetition rates up to \unit[1]{kHz}, although guiding of high-intensity laser pulses has not yet been demonstrated at this repetition rate \cite{Gonsalves:2016jc}.

In hydrodynamically formed plasma channels \cite{Durfee:1993, Durfee:1994wz, Durfee:1995gr, Clark:1997we, Clark:2000dk} a cylindrical region of plasma is formed, then heated, by one or more laser pulses. Rapid  expansion of the plasma column drives a radial shock into the cold, unionized ambient gas to form a transverse electron density profile which increases with radial distance out to the position of the shock front. Channels of this type are a promising solution for practical plasma accelerators capable of operation at high (multi-kilohertz) repetition rates, since the plasma channel is free-standing, with no nearby physical structure which could be damaged by the driving laser pulse.

To date the initial plasma column in hydrodynamic channels has been heated by laser-driven electron-ion collisions. However, since rapid collisional heating requires high plasma densities it has proved difficult to generate channels with low on-axis densities. For example, Milchberg et al have demonstrated the generation of channels with  $n_{e}(0) \approx \unit[5 \times 10^{18}]{cm^{-3}}$ in argon gas \cite{Clark:1997we}. By exploiting the fact that in a clustered gas the local atomic density is much higher than the mean density, the same group demonstrated the generation of channels with $n_{e}(0) \approx \unit[1 \times 10^{18}]{cm^{-3}}$ channels in clustered argon and hydrogen targets \cite{Kumarappan:2005du}.

In this paper we describe the generation of plasma channels by hydrodynamic expansion of a plasma column formed and heated by optical field ionization (OFI) with elliptically-polarized laser pulses. The mean energy of electrons ionized by this mechanism can be controlled by adjusting the ellipticity of the laser field; further, since OFI operates at the atomic level, the electron heating is independent of the initial density, which allows the formation of low density plasma channels. We employ simulations and experiments to demonstrate the formation of plasma channels with axial densities in the range $\unit[1.5 \times 10^{17}]{cm^{-3}} \lesssim n_\mathrm{e}(0) \lesssim \unit[1 \times 10^{18}]{cm^{-3}}$, i.e.\ an axial density an order of magnitude lower than achieved to date. The  matched spot sizes of the lowest-order-modes are found to be in the range $\unit[61]{\mu m} \gtrsim W_\mathrm{M} \gtrsim  \unit[33]{\mu m}$, and the $1/\mathrm{e}$ intensity attenuation lengths are of order \unit[160]{mm}.  Simulations show that the laser energy required to generate these channels is of order \unit[1]{mJ} per centimetre of channel. 

We note that previous authors \cite{Lemos:2013gb, Lemos:2013ju} have demonstrated the formation of short (\unit[4]{mm} long) plasma channels by OFI of hydrogen and helium. However, the channels demonstrated in that work had high axial electron densities of $n_\mathrm{e}(0) \gtrsim \unit[1 \times 10^{18}]{cm^{-3}}$, and the authors do not appear to have considered the use of this approach to generate long, low density channels, as we propose and demonstrate here.

The paper is organized as follows. We first describe analytic and numerical models of the formation of the plasma column and its subsequent expansion. Section \ref{Sec:Experiments} describes experiments to characterize the properties of short, low-density plasma channels generated in hydrogen by focusing femtosecond-duration laser pulses with a conventional lens. Section \ref{Sec:Axicon} presents numerical modelling of long, low-density plasma channels produced by an axicon lens, and in Section \ref{Sec:Conclusion} we draw conclusions.

\section{Simulations of OFI-driven hydrodynamic channels}
\subsection{Electron Energies from OFI Heating} \label{sec: Electron Energies from OFI Heating}
Optical field ionization (OFI) occurs when the strength of an applied electric field becomes comparable to the fields binding the valence electrons in the target atom.  The subsequent motion of the ionized electron has two components: a driven oscillation at the laser frequency; and a constant drift \cite{Corkum:1989ji, Gallagher:1988er}. After the laser pulse has passed, only the drift component remains, with momentum $\vec{p}_\mathrm{f} = e \vec{A}(t_0)$ where $\vec{A}(t)$ is the vector potential of the laser field and the electron is ionized at time $t = t_0$.

The electrons will predominantly be ionized when the magnitude of the laser field $\vec{E}(t) = - \partial \vec{A} / \partial t$ is close to a maximum. Hence, for linearly polarized radiation $\vec{A}(t_0) \approx 0$ and the retained momentum will be small. In contrast, for circular polarization the drift energy will be large. To see this, suppose that the electron is ionized when the field points along the $x$-axis, and hence $A_x(t_0) =0$. For laser propagation along the $z$-axis then $\left|A_y(t_0)\right| = E_0 / \omega$, where $E_0$ is the magnitude of the electric field. Hence the electrons will retain a momentum $p_y = e E_0/\omega$ and a kinetic energy $E^\mathrm{circ}_\mathrm{k} = e^2 E_0^2 / 2 m_\mathrm{e} \omega^2 = U_\mathrm{p}(t_0)$, where $U_\mathrm{p}(t_0)$ is the ponderomotive energy of the laser field at the moment of ionization.

Optical field ionization therefore produces electrons with momentum in the plane transverse to the laser propagation; the magnitude of this momentum will increase as the ellipticity of the field $\epsilon$ (defined as the ratio of the minor to major axes of the ellipse) is increased towards unity.

\begin{center}
\begin{figure}[tb]
\includegraphics[trim=0cm 0 0 0]{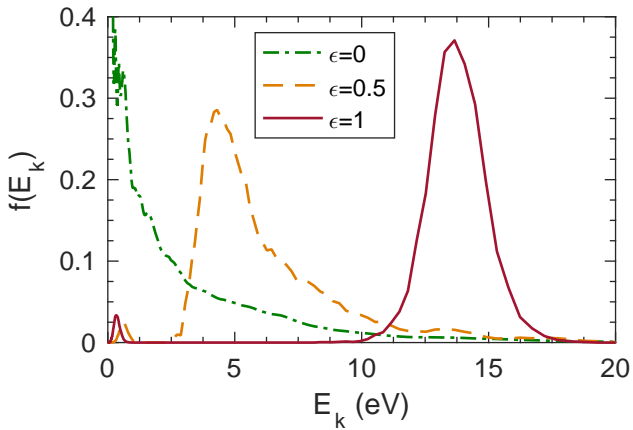}
\caption{Calculated retained kinetic energy distributions $f(E_\mathrm{k})$ of electrons produced by optical field ionization of molecular hydrogen with laser fields of various ellipticity $\epsilon$. For these simulations the laser pulse was assumed to have a Gaussian temporal profile of \unit[40]{fs} full-width at half maximum (FWHM), a peak intensity $\unit[2.5 \times 10^{14}]{W cm^{-2}}$, and a centre wavelength of $\lambda = \unit[800]{nm}$.}
\label{Fig:H_energy_spectra}
\end{figure}
\end{center}

Figure \ref{Fig:H_energy_spectra} shows, for three different laser ellipticities, the electron energy distribution following ionization of molecular hydrogen calculated by the particle-in-cell (PIC) code EPOCH \cite{Arber2015}. The EPOCH code calculates the ionization rates using the ADK model in the tunnelling regime, and the Posthumus model \cite{Posthumus1997} in the BSI regime \cite{Lawrence-Douglas2013}, and it subsequently tracks particle motion in the electric field. For these simulations the laser pulse was assumed to have a Gaussian temporal profile of \unit[40]{fs} full-width at half maximum (FWHM), a peak intensity of $\unit[2.5 \times 10^{14}]{W cm^{-2}}$, and a centre wavelength of $\lambda = \unit[800]{nm}$. The optical field ionization of hydrogen is complicated by the fact that it is a diatomic molecule and hence there are several paths which result in fully ionized, dissociated hydrogen atoms \cite{Tauscher:2017}. For simplicity we have treated the hydrogen as being atomic, but with an ionization energy of \unit[15.4]{eV}, corresponding to the first ionization of \emph{molecular} hydrogen \cite{Tauscher:2017, Shiner:1993}. A more advanced treatment would be required to account for the multiple dissociation and ionization pathways, but the electron energy distribution is not expected to be very different from that calculated here. As expected, the mean electron energy increases with the ellipticity: from $\left< E_k \right> = \unit[1.8]{eV}$ for $\epsilon = 0$ (linear polarization) to $\left< E_k \right> = \unit[13.7]{eV}$ for $\epsilon = 1$ (circular polarization).

\subsection{Modelling Shock Expansion} \label{Sec:Modelling_shock_expansion}
The non-isotropic and non-thermal electron momentum distribution complicates a calculation of the subsequent expansion of plasma columns produced by OFI. For the general case of multi-electron ionization to form an ion of atomic charge $Z > 1$, ionization of each valence electron gives rise to a ``class'' of electrons with a mean energy determined by its ionization energy. For each class the momentum distribution will become isotropic, and the energy distribution will thermalize, in a time of order the Spitzer electron self collision time   $\tau^\mathrm{e}_\mathrm{c}  \approx (1.40/ 8 \pi r_\mathrm{e}^2 c^4 n_\mathrm{e} \ln \Lambda) (3 k_\mathrm{B} T_\mathrm{e}/m_\mathrm{e})^{3/2} $, where $\ln \Lambda$ is the Coulomb logarithm, and $T_\mathrm{e}$ is the effective temperature  of the class \cite{Spitzer1967}. For $T_\mathrm{e} = \unit[10]{eV}$ and $n_\mathrm{e} = \unit[10^{18}]{cm^{-3}}$, we find $\tau^\mathrm{e}_\mathrm{c}  \approx \unit[0.9]{ps}$. The different electron classes will exchange energy, to form a single thermal distribution, in a time of order $\tau^\mathrm{e}_\mathrm{c}$ calculated with $T_\mathrm{e}$ equal to the mean temperature of all the classes; this timescale is also around \unit[1]{ps}.

The hot electrons produced by OFI will start to stream out of the initial plasma column but  will be held back by the cold, positive ions. If a fraction $\alpha$ of high energy electrons have already escaped a plasma column of radius $r_\mathrm{c}$, only electrons with kinetic energy of $E_\mathrm{k} > \alpha m_e \omega_\mathrm{p}^2 r_c^2$ can escape the cylinder. Assuming $r_\mathrm{c} \approx \unit[10]{\mu m}$, and $n_e \sim \unit[10^{18}]{ cm^{-3}}$, this condition yields $\alpha \lesssim 10^{-6} E_\mathrm{k} \mathrm{[eV]}$. The Debye length $\lambda_\mathrm{D} = \sqrt{{\epsilon_0 k_B T} /{n_e e^2}}$ is found to be $\sim \unit[10]{nm}$, and hence we expect only a very small proportion of electrons to stream out of the expanding plasma, and for the plasma to remain locally neutral.

The considerations above show that after a time of order a few ps the electron distribution will be isotropic and thermal, with a temperature $k_\mathrm{B} T_\mathrm{e} = (2/3) \left< E_k \right>$, where $\left< E_k \right>$ is the mean electron kinetic energy immediately after OFI. The ions will remain cold. The subsequent expansion of the plasma can therefore be modelled by fluid codes.

We have modelled the hydrodynamic expansion of OFI plasma columns using two different Lagrangian single fluid codes: HELIOS \cite{MacFarlane2006}, and an in-house code developed by one of the authors \cite{Durfee:1993, Durfee:1995gr, Fan:2000fh}. Both codes assume Maxwellian energy distributions for electrons and ions, but do not assume $T_\mathrm{e} = T_\mathrm{ion}$.

The in-house code includes a model of OFI and solves the Helmholtz wave equation for the propagation of the channel-forming pulse in the evolving plasma \cite{Fan:2000fh}. The channel-forming pulse is assumed to be focused by an axicon lens, and the plasma expansion is subsequently calculated using the single-fluid equations for mass, momentum and energy conservation. In addition to modelling OFI, the code includes inverse-bremsstrahlung heating, collisional ionization and recombination, and thermal conduction.

The HELIOS code does not include OFI and provides only an opacity model for energy deposition. Therefore for the HELIOS simulations the initial electron density profile simulated by the in-house code was used to describe the initial fraction of ionized atoms; the initial electron temperature was calculated from the electron energy spectrum shown in Figure~\ref{Fig:H_energy_spectra}. The ion temperature, and the temperature of the neutral gas, were assumed to be $T_\mathrm{ion} = \unit[298]{K}$ everywhere.

A useful analytic expression for the temporal evolution of the radial position of the shock front is provided by the Sedov-Taylor solution for expansion of ideal gases, assuming $T_\mathrm{e} = T_\mathrm{ion}$ \cite{Taylor1950,Hutchens1995}. For an idealized system comprising an initial cylindrical region of infinitesimal radius and energy per unit length $E_\sigma$, expanding into an unshocked region of density $\rho_0$, the radial position of the shock front is given by,

\begin{align}
r_\mathrm{s}^4 = \frac{(\gamma+1)^2}{\pi}\frac{E_\sigma \tau^2}{\rho_0},
\end{align}
where $\tau=0$ is the idealized moment when $r_\mathrm{s} = 0$. When fitting to experimental data we will write $\tau=t + \tau_0$, where $t$ is the delay after the initial deposition of energy along the axis, and the constant $\tau_0 > 0$. In our case $E_\sigma = \left(Z n_0 E_\mathrm{k} \right) \pi r_0^2$, where $r_0$ is the radius of the initial plasma column and $n_0$ is the initial ion density, and $\rho_0 = n_0 M_\mathrm{ion}$. Hence the Sedov-Taylor solution gives,

\begin{align}
r_\mathrm{s}(t) =  \left(\gamma+1\right)^{1/2} \left(\frac{ Z E_\mathrm{k}}{M_\mathrm{ion}} \right)^{1/4} \left[r_0(t+\tau_0)\right]^{1/2},\label{Eqn:Sedov-Taylor}
\end{align}
where, for a fully ionized medium, the adiabatic index $\gamma = 5/3$.

In the section below we compare the results of these models with the results of experiments on the hydrodynamic expansion of hydrogen plasma produced by OFI.

\section{Experimental demonstration of channel formation}\label{Sec:Experiments}
\begin{figure}[tb]
\begin{center}
\includegraphics[width=9cm]{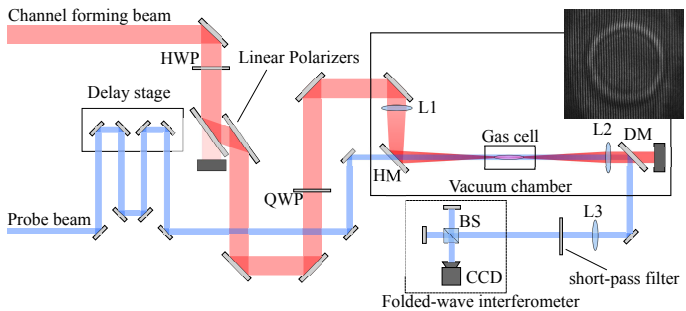}   
\end{center}
\caption{Schematic diagram of the experimental layout employed to measure low density plasma channels. Symbols: Half-wave plate (HWP); quarter-wave plate (QWP); plano-convex lenses ($L1$ - $L3$); dichroic mirror ($DM$); 50/50 beamsplitter (BS). The inset shows the interference fringes observed when a plasma channel is formed.}
\label{Fig:ExpLayout}
\end{figure}

To demonstrate the potential of this approach for generating low density plasma channels, we generated short plasma channels by focusing femtosecond-duration laser pulses with a conventional lens. Figure \ref{Fig:ExpLayout} shows the experimental arrangement employed.

The channels were formed by pulses from a Ti:sapphire laser system of central wavelength $\lambda = \unit[800]{nm}$ and \unit[50]{fs} FWHM duration. This channel-forming beam was passed through a quarter-wave plate which converted the beam from linear to elliptical polarization, with an ellipticity which could be varied by adjusting the angle between the incident polarization and the fast axis of the wave plate. This beam was then directed into a vacuum chamber, focused by a  fused-silica plano-convex lens of focal length $f = \unit[500]{mm}$, used at $f/13$, and directed into a gas cell by a mirror (HM) in which a \unit[4]{mm} hole was drilled at $45^{\circ}$ to the normal of the mirror face. A combination of a half-wave plate and linear polarizer prior to the quarter-wave plate allowed the energy of the pulses reaching the gas cell to be adjusted.

A synchronized \unit[400]{nm}, \unit[3]{mm} diameter probe beam was formed by frequency doubling a small fraction of the \unit[800]{nm} beam. The probe beam was: directed to a four-pass, motorized delay stage to control the delay $t$ between the arrival of the channel-forming and probe pulses; injected into the gas cell, by passing it through the hole in HM; and aligned to be collinear with the channel-forming beam by mirrors external to the vacuum chamber.

The variable-length gas cell was machined from aluminium with a single gas inlet fed from a reservoir placed outside the vacuum chamber. Hydrogen gas could be flowed into the cell via a \unit[4]{mm} inner diameter pipe and the pressure within the cell was measured by a capacitance manometer (MKS Instruments Baratron 626) placed \unit[1]{m} from the cell. The gas input was pulsed using a solenoid valve to limit the background chamber pressure. The channel-forming and probe beams were coupled into and out of the gas cell  through $\unit[750]{\mu m}$ diameter pinholes drilled in $\unit[200]{\mu m}$ thick stainless steel sheet and spaced by either \unit[2]{mm} or \unit[4]{mm}. The length of the gas cell was measured prior to the experiments, and the value confirmed by measuring, as a function of fill pressure, the probe phase shift immediately after the arrival of the channel-forming pulse.

After propagating through the gas cell the channel-forming and probe beams were separated by a dichroic mirror ($DM$, \unit[400]{nm} reflecting, \unit[800]{nm} transmitting). The diameter of the probe beam was increased by a factor of $\sim3.7$ by a Keplerian telescope formed by a pair of plano-convex lenses ($L2$ and $L3$) of focal lengths $f_2 = \unit[270]{mm}$ and $f_3 = \unit[1000]{mm}$, and directed to a folded-wave interferometer; this comprised a Michelson interferometer, adjusted so that the two exiting beams formed straight, non-localized fringes with a spacing which could be varied by adjusting its mirrors (see inset to Fig.\ \ref{Fig:ExpLayout}).

The front focal plane of $L2$ was adjusted to coincide with the exit pinhole of the gas cell, and an 8-bit CCD camera was positioned in the back focal plane of $L3$ so that it imaged the exit pinhole of the gas cell. In separate calibration experiments the resolution of the imaging system was found to be $\unit[(0.83 \pm 0.01)]{\mu m /pixel}$.

In the presence of plasma, those parts of the probe beam passing through the plasma acquired an additional phase $\phi(x,y) = n_\mathrm{e}(x,y) r_\mathrm{e} \lambda_\mathrm{probe} \ell$, where $x$ and $y$ are transverse coordinates measured from the axis of the initial plasma column, $\lambda_\mathrm{probe}$ is the probe wavelength, and  $\ell$ is the length of the plasma column.  This additional phase shift, and hence the transverse electron density profile, $n_\mathrm{e}(x,y)$, could be extracted from the  interferogram by standard methods \cite{Takeda:1982,Bone:1986}. Beam propagation simulations, using the extracted electron density profiles, showed that refraction of the probe beam by the plasma was not significant for the conditions of these experiments.

\begin{figure}[tb]
\centering
\includegraphics[width=\columnwidth]{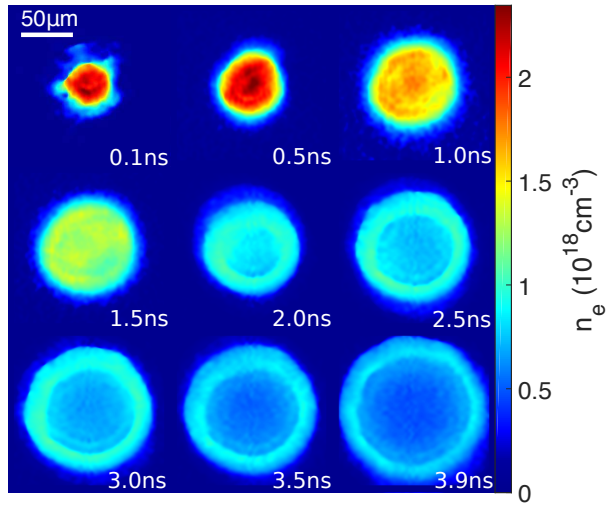}
\caption{Measured transverse electron density profiles at delays $t$ after the arrival of the channel-forming pulse focused into \unit[50]{mbar} of hydrogen gas. Each profile is obtained from analysis of a single interferogram and is shown in a square of side $\unit[280]{\mu m}$. The delay $t$ is indicated for each plot.}
\label{Fig:ElectronDensityProfiles}
\end{figure}

Figure \ref{Fig:ElectronDensityProfiles} shows the extracted electron density profiles measured for a circularly polarised channel-forming pulse of energy $\unit[(26.7 \pm 2.9)]{mJ}$ and an initial cell pressure of \unit[50]{mbar}. The initial plasma formed ($t = 0$) comprises an approximately cylindrical region of diameter $\unit[72]{\mu m}$ and peak electron density $n_\mathrm{e}(0) \approx \unit[2.4\times10^{18}]{cm^{-3}}$, corresponding to full ionization of the hydrogen gas. The subsequent hydrodynamic expansion of the plasma column to form a plasma channel is clearly evident.

\begin{figure}[tb]
    \centering
    \includegraphics[width=\columnwidth]{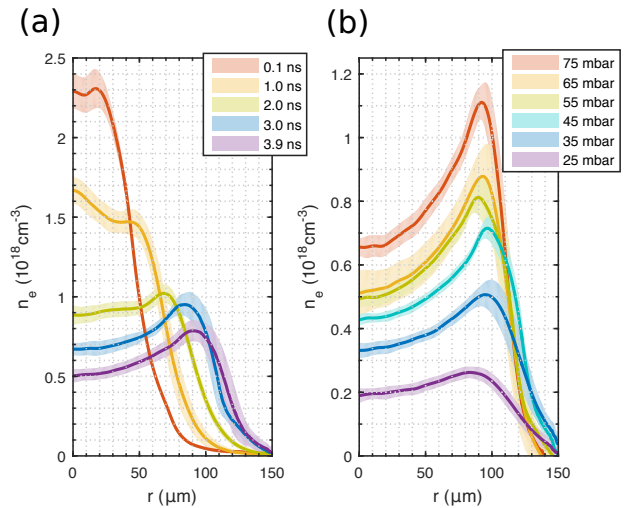}
    \caption{Measured transverse electron density profile for hydrogen plasma after OFI by a circularly polarized pulse. The solid lines show the mean rotationally averaged profile for 5-10 shots while the associated coloured band shows the RMS error in the measurement. (a) shows the temporal evolution of a plasma channel from an initial fill pressure of \unit[50]{mbar} while (b) shows the electron density profile \unit[3.9]{ns} after ionization for a range of different cell fill pressures.}
    \label{Fig:Measured_channel_evolution}
\end{figure}

Figure \ref{Fig:Measured_channel_evolution}  shows measured transverse electron density profiles for the same laser parameters as in Fig.\ \ref{Fig:ElectronDensityProfiles}.  For each shot, the transverse electron density profile $n_\mathrm{e}(r)$, where $r^2 = x^2 + y^2$, was found by rotationally averaging about the axis --- the position of the axis being determined by fitting an ellipse to the FWHM contour. To reduce the effect of the large shot-to-shot jitter in the pulse energy of the channel forming beam (approximately 10--15\% rms in these experiments), the electron density profile was averaged over 5-10 shots.  Figure \ref{Fig:Measured_channel_evolution} (a) shows the temporal evolution of the channels formed for a cell pressure of \unit[50]{mbar}.  For this initial pressure, at $t = \unit[3.9]{ns}$ the plasma column has driven a cylindrical shock wave, peaked at $r \approx \unit[92]{\mu m}$; the electron density increases from $n_\mathrm{e}(0) \approx \unit[5.0\times10^{17}]{cm^{-3}}$ to approximately $\unit[7.9\times10^{17}]{cm^{-3}}$ at the peak of the shock front.

Figure \ref{Fig:Measured_channel_evolution} (b) shows the variation with the initial cell pressure of transverse electron density profile at a fixed delay of $t = \unit[3.9]{ns}$. A plasma channel is formed in all cases, with an axial density which is  proportional to the initial cell pressure within experimental error. For an initial cell pressure $P = \unit[25]{mbar}$ the axial density is $n_\mathrm{e} (0) \approx \unit[(1.9 \pm 0.2) \times 10^{17}]{cm^{-3}}$.

The rotationally-averaged electron density profiles at $t = \unit[3.9]{ns}$ had axial densities in the range $n_\mathrm{e}(0) = \unit[1.5\times10^{17}]{cm^{-3}}$ to $\unit[1.0\times10^{18}]{cm^{-3}}$; these profiles were used to calculate the lowest-order modes of the channels by solving the Helmholtz equation for the electric field of the form $E(\vec{r},z) = u(\vec{r})\exp(i \beta z)$, where $z$ is the position along the waveguide axis and $\vec{r}$ is the perpendicular position vector \cite{Clark:2000dk}. The $1/\mathrm{e}$ attenuation length for the \emph{power} of the guided mode is then $L_\mathrm{attn} = [2 \Im(\beta)]^{-1}$.

For $P = \unit[25]{mbar}$ ($n_\mathrm{e}(0) \approx \unit[1.9\times10^{17}]{cm^{-3}}$ at $t = \unit[3.9]{ns}$) the measured channel has a calculated lowest-order mode with a matched spot size (defined as the radius at which the intensity is a factor $1/e^2$ smaller than that on axis) of  $W_\mathrm{M} = \unit[48]{\mu m}$, and an attenuation length of $L_\mathrm{attn} = \unit[160]{mm}$.

\begin{center}
\begin{figure}[tb]
\includegraphics[trim=0.5cm 0 0 0]{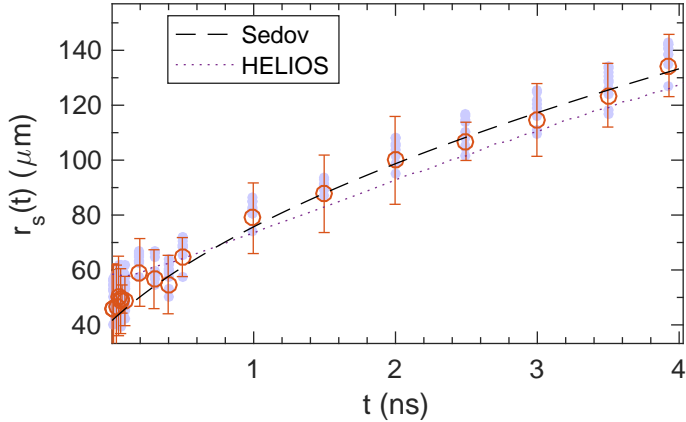}
\caption{Comparison of measured and simulated temporal evolution of the shock front. Blue circles show the average of the major and minor axes of an ellipse fitted to a half-max contour of the measured electron density profiles; the same data binned by time interval is shown as open red circles. The results of a HELIOS simulation (dotted purple) and a fit of the data to the Sedov-Taylor solution (eqn \eqref{Eqn:Sedov-Taylor}), dashed black) are also shown.}
\label{Fig:Shock_front_vs_time}
\end{figure}
\end{center}

Figure \ref{Fig:Shock_front_vs_time} compares the measured expansion of the plasma channel with the HELIOS simulations and the Sedov-Taylor theory described in \ref{Sec:Modelling_shock_expansion}.  For both measured and simulated electron density profiles the shock radius was taken to be the half-width at half-maximum (HWHM) of the electron density profile. For the Sedov-Taylor solution the initial hot electron energy was set equal to that calculated from the EPOCH simulations shown in Fig.~\ref{Fig:H_energy_spectra} (i.e.\ $\left< E_\mathrm{k} \right> = \unit[13.7]{eV}$). The equivalent initial plasma radius extracted from a fit of eqn\ \eqref{Eqn:Sedov-Taylor} to the data is $r_0 = \unit[(41.5 \pm 1.5)]{\mu m}$, which is consistent with the measured data as $t \rightarrow 0$. The HELIOS simulation assumed an initial electron temperature $k_\mathrm{B} T_\mathrm{e} = (2/3) \left< E_\mathrm{k} \right> = \unit[9.1]{eV}$ and an initial transverse electron density profile described by a 10th order super-Gaussian chosen to match the measured profile of the initial plasma column. It can be seen that both the HELIOS simulations and the Sedov-Taylor solution are in excellent agreement with the experimental data, demonstrating that  the essential physics is captured by our models of the OFI heating of the initial plasma column and its subsequent expansion.

\begin{center}
\begin{figure}[tb]
\includegraphics[trim=0.5cm 0 0 0] {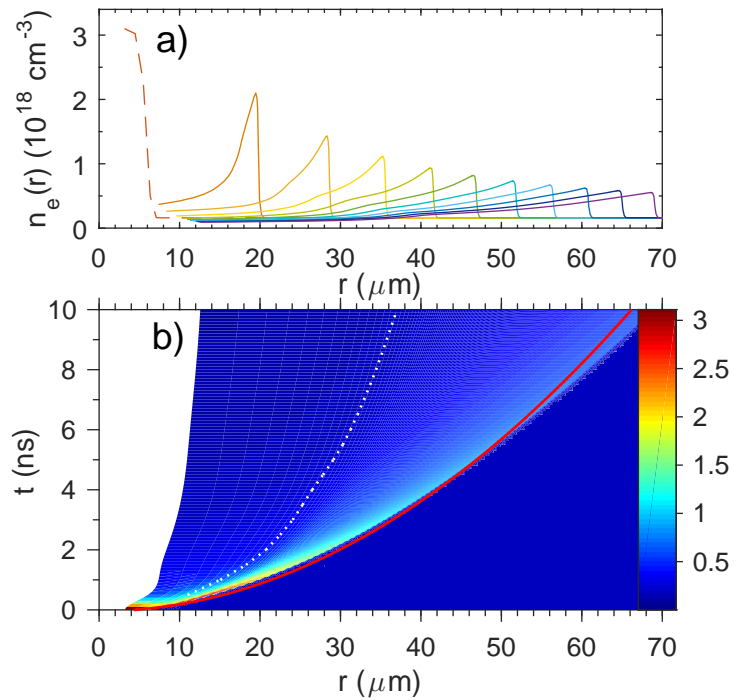}
\caption{Simulated evolution of the transverse density profile $n_\mathrm{e}(r)$ of the hydrodynamic expansion of an OFI plasma produced in molecular hydrogen of pressure \unit[60]{mbar} by an axicon  with $\alpha = 2.5^\circ$ and $R = \unit[18]{mm}$. (a) $n_\mathrm{e}(r)$ at \unit[1]{ns} intervals after the  formation of the initial plasma column (dashed line). (b) The calculated electron density surface, overlaid with the shock front position $r_s(t)$ expected from the Sedov-Taylor solution (red) and the calculated matched spot size $W_\mathrm{M}$ of the lowest-order mode of the plasma channel (dotted white). For these simulations the incident channel-forming pulses were assumed to be circularly-polarized, of top-hat incident transverse profile, of energy \unit[22]{mJ}, and \unit[40]{fs} FWHM duration.}
\label{Fig:Axicon_Sim_Composite}
\end{figure}
\end{center}

\section{Long Plasma Channels produced by axicon foci}\label{Sec:Axicon} 
The  results presented above demonstrate that short, low-density plasma channels can be generated by hydrodynamic expansion of an OFI plasma, and that the properties of those channels are in excellent agreement with our analytic and numerical models. We now consider the extension of this approach to the generation of long, low-density plasma channels produced by expansion of plasma columns formed in a longitudinally extended focal region such as that produced by an axicon or axilens \cite{McLeod:1954, Sochacki:1992}.

We will concentrate on the use of an axicon lens, as successfully used in collisionally-heated hydrodynamic channels \cite{Durfee:1993, Durfee:1994wz, Durfee:1995gr, Clark:1997we}.  The length  of the axicon focus is given by $L = R \cot \alpha$, where $R$ is the radius of the axicon and $\alpha$ is the approach angle, i.e.\ the angle between rays refracted by the axicon and the axis. In vacuo, the intensity of the beam downstream of the axicon is given by $I(r, z) = A(z) J_0^2(k r \sin \alpha)$, where $k = 2 \pi /\lambda$, and $A(z)$ depends on the axicon properties and the input power and transverse intensity profile of the incident laser radiation. Hence the transverse profile of the beam is strongly peaked on axis, with a first zero at a radial distance $r_1 = b_1 / 2 \pi \sin \alpha$, where $b_1 \approx 2.4$ is the position of the first zero of the zeroth-order Bessel function. The presence of plasma introduces a further constraint in that refraction will cause rays  approaching the axis to be reflected at a modified critical density $n_\mathrm{crit}(\alpha) = n_\mathrm{c} \sin^2 \alpha$, where $n_\mathrm{c} = 1/ r_\mathrm{e} \lambda^2$ is the critical density at normal incidence. If we neglect the effects of evanescent waves, this effect places an upper limit on the electron density which can be generated. 

Figure \ref{Fig:Axicon_Sim_Composite} shows simulations of the formation of a \unit[200]{mm}-long plasma channel by circularly-polarized $\lambda = \unit[800]{nm}$ laser pulses of top-hat incident transverse profile, \unit[22]{mJ} energy, with a Gaussian temporal profile of \unit[40]{fs} FWHM duration. The axicon parameters were taken to be $\alpha = 2.5^\circ$ and $R = \unit[18]{mm}$, and the initial gas was assumed to be molecular hydrogen with a pressure of \unit[60]{mbar} and a temperature of \unit[298]{K}; under these conditions the initial electron density was close to $n_\mathrm{crit}(\alpha)$. The initial transverse electron density profile calculated by the in-house code was approximated by a 6th order super-Gaussian of radius $\unit[5]{\mu m}$, which in turn was used as the initial electron density profile for a HELIOS simulation. The energy of the ionized electrons was set to $E_k=\unit[13.7]{eV}$, calculated from the energy distributions shown in Fig.\ \ref{Fig:H_energy_spectra}. It is seen that the initial plasma column expands in a time of a few nanoseconds to form a plasma channel; the on-axis density of the channel is found to decrease from $n_\mathrm{e}(0) \approx \unit[4 \times 10^{17}]{cm^{-3}}$ at $ t = \unit[1]{ns}$ to $n_\mathrm{e}(0) \approx \unit[0.9 \times 10^{17}]{cm^{-3}}$ at $ t = \unit[10]{ns}$. The position of the shock front is found to agree well with the Sedov-Taylor solution, where the effective hard edge plasma column radius was fitted as $r_0=\unit[4.5]{\mu m}$.

\begin{center}
\begin{figure}[tb]
\includegraphics[trim=0.5cm 0 0 0] {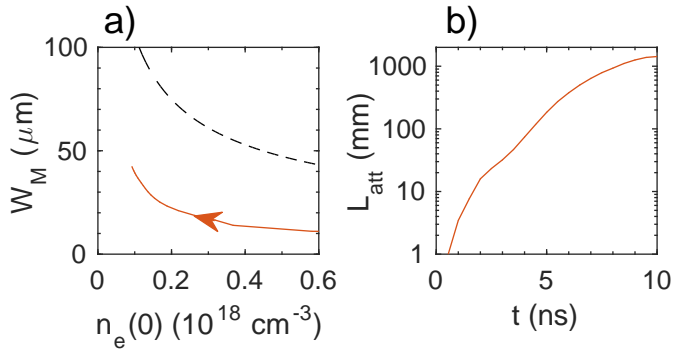}
\caption{(a) The spot size (solid red line) $W_\mathrm{M}$ of the lowest-order mode plotted against the axial electron density $n_\mathrm{e}(0)$ of the plasma channel for the plasma channel shown in Fig.\ \ref{Fig:Axicon_Sim_Composite} during the interval $0 \leq t \leq \unit[10] ns$. The channel starts at the bottom right of the plot and evolves towards the top left, corresponding to a decreasing on-axis density and an increasing matched spot size. Also shown is the evolution of $\lambda_\mathrm{p}$ (dashed black); to remain in the quasi-linear regime with $a_0 \approx 1$ and $P < P_\mathrm{c}$, a guided pulse should have a spot size less than this value. (b) Temporal evolution of the 1/e power attenuation length $L_\mathrm{att}$ of the lowest-order mode.}
\label{Fig:Axicon_Channel_Properties}
\end{figure}
\end{center}

The modes of the simulated axicon-generated plasma channels were calculated at different times during the expansion, using the in-house Helmholtz code \cite{Clark:2000dk}. Figure \ref{Fig:Axicon_Sim_Composite} also shows the evolution of the matched spot size $W_\mathrm{M}$ of the lowest-order mode as the channel expands; it can be seen that in this case the spot size increases from approximately $\unit[10]{\mu m}$ at early times to $\unit[40]{\mu m}$ for $t \approx \unit[10]{ns}$. Figure \ref{Fig:Axicon_Channel_Properties}(a) plots the matched spot size of the plasma channel  against its axial density during the expansion. It can be seen that the spot size remains below $\lambda_\mathrm{p}$, as required to remain in the qausi-linear regime whilst ensuring that the peak power of the guided pulse $P < P_\mathrm{c}$. Figure \ref{Fig:Axicon_Channel_Properties}(b) shows the temporal evolution of the power attenuation length of the lowest-order mode; from this it can be seen that for $t \gtrsim \unit[5]{ns}$ low-loss guiding is possible over hundreds of millimetres.

\section{Conclusion}\label{Sec:Conclusion}
We have proposed that the hydrodynamic expansion of columns of plasma produced by optical field ionization could generate long plasma channels with low propagation losses. Since OFI operates at the atomic level, the electron heating is independent of the initial density, which allows the formation of low density plasma channels. 

These ideas have been confirmed by experiments with a conventional lens which show the formation of short plasma channels with on-axis densities $\unit[1.5 \times 10^{17}]{cm^{-3}} \lesssim n_\mathrm{e}(0) \lesssim \unit[1 \times 10^{18}]{cm^{-3}}$. These measured plasma channels could support lowest-order modes of spot size $\unit[61]{\mu m} \gtrsim W_\mathrm{M} \gtrsim  \unit[33]{\mu m}$ and $1/\mathrm{e}$ power attenuation lengths of order \unit[100]{mm}. Furthermore, numerical simulations demonstrate that an axicon lens could be used to generate long plasma channels with on-axis densities on the order of $n_e \approx \unit[10^{17}]{cm^{-3}}$, matched spot sizes around $W_\mathrm{M}\approx \unit[40]{\mu m}$, and attenuation lengths of order $L_\mathrm{att} \sim \unit[1000]{mm}$. 

Hydrodynamic OFI (HOFI) channels of this type could be generated repeatedly, and, since they are free-standing, the waveguide system would be immune to damage or degradation by the guided laser pulse or pulses. The maximum possible repetition rate would be limited by the time for the plasma to recombine and return to uniform density, or by the time taken to sweep fresh gas into the channel region. Assuming a flow velocity of order $\unit[1]{km s^{-1}}$, and a transverse scale of $\unit[100]{\mu m}$, repetition rates up to the $\unit[]{MHz}$ range would seem to be possible in principle. We note that for the axicon-generated channels shown in Figs \ref{Fig:Axicon_Sim_Composite} and \ref{Fig:Axicon_Channel_Properties}, the required laser energy was only approximately \unit[1]{mJ} per centimetre of channel.

The prospect of generating low density channels with lengths of order \unit[1]{m} at pulse repetition rates of at least several kilohertz would therefore appear to be realistic. HOFI channels of this type would appear to be an ideal basis for multi-GeV laser-plasma accelerator stages capable of long-term operation at high pulse repetition rates.

This work was supported by the UK Science and Technology Facilities Council (STFC UK) [grant numbers ST/J002011/1, ST/M50371X/1,  ST/N504233/1, and ST/P002048/1]; the Helmholtz Association of German Research Centres [Grant number VH- VI-503]; and Air Force Office of Scientific Research, Air Force Material Command, USAF [Grant number FA8655-13-1-2141]. HMM acknowledges the support of the US Dept. of Energy (DESC0015516), the Dept. of Homeland Security (2016DN077ARI104), the National Science Foundation (PHY1619582), and the AFOSR (FA95501610284).

\bibliography{References}

\begin{thebibliography}{56}%
\makeatletter
\providecommand \@ifxundefined [1]{%
 \@ifx{#1\undefined}
}%
\providecommand \@ifnum [1]{%
 \ifnum #1\expandafter \@firstoftwo
 \else \expandafter \@secondoftwo
 \fi
}%
\providecommand \@ifx [1]{%
 \ifx #1\expandafter \@firstoftwo
 \else \expandafter \@secondoftwo
 \fi
}%
\providecommand \natexlab [1]{#1}%
\providecommand \enquote  [1]{``#1''}%
\providecommand \bibnamefont  [1]{#1}%
\providecommand \bibfnamefont [1]{#1}%
\providecommand \citenamefont [1]{#1}%
\providecommand \href@noop [0]{\@secondoftwo}%
\providecommand \href [0]{\begingroup \@sanitize@url \@href}%
\providecommand \@href[1]{\@@startlink{#1}\@@href}%
\providecommand \@@href[1]{\endgroup#1\@@endlink}%
\providecommand \@sanitize@url [0]{\catcode `\\12\catcode `\$12\catcode
  `\&12\catcode `\#12\catcode `\^12\catcode `\_12\catcode `\%12\relax}%
\providecommand \@@startlink[1]{}%
\providecommand \@@endlink[0]{}%
\providecommand \url  [0]{\begingroup\@sanitize@url \@url }%
\providecommand \@url [1]{\endgroup\@href {#1}{\urlprefix }}%
\providecommand \urlprefix  [0]{URL }%
\providecommand \Eprint [0]{\href }%
\providecommand \doibase [0]{http://dx.doi.org/}%
\providecommand \selectlanguage [0]{\@gobble}%
\providecommand \bibinfo  [0]{\@secondoftwo}%
\providecommand \bibfield  [0]{\@secondoftwo}%
\providecommand \translation [1]{[#1]}%
\providecommand \BibitemOpen [0]{}%
\providecommand \bibitemStop [0]{}%
\providecommand \bibitemNoStop [0]{.\EOS\space}%
\providecommand \EOS [0]{\spacefactor3000\relax}%
\providecommand \BibitemShut  [1]{\csname bibitem#1\endcsname}%
\let\auto@bib@innerbib\@empty
\bibitem [{\citenamefont {Hooker}(2013)}]{Hooker:2013jk}%
  \BibitemOpen
  \bibfield  {author} {\bibinfo {author} {\bibfnamefont {S.~M.}\ \bibnamefont
  {Hooker}},\ }\href {\doibase 10.1038/nphoton.2013.234} {\bibfield  {journal}
  {\bibinfo  {journal} {Nature Photon}\ }\textbf {\bibinfo {volume} {7}},\
  \bibinfo {pages} {775} (\bibinfo {year} {2013})}\BibitemShut {NoStop}%
\bibitem [{\citenamefont {Leemans}\ \emph {et~al.}(2006)\citenamefont
  {Leemans}, \citenamefont {Nagler}, \citenamefont {Gonsalves}, \citenamefont
  {Toth}, \citenamefont {Nakamura}, \citenamefont {Geddes}, \citenamefont
  {Esarey}, \citenamefont {Schroeder},\ and\ \citenamefont
  {Hooker}}]{Leemans:2006}%
  \BibitemOpen
  \bibfield  {author} {\bibinfo {author} {\bibfnamefont {W.~P.}\ \bibnamefont
  {Leemans}}, \bibinfo {author} {\bibfnamefont {B.}~\bibnamefont {Nagler}},
  \bibinfo {author} {\bibfnamefont {A.~J.}\ \bibnamefont {Gonsalves}}, \bibinfo
  {author} {\bibfnamefont {C.}~\bibnamefont {Toth}}, \bibinfo {author}
  {\bibfnamefont {K.}~\bibnamefont {Nakamura}}, \bibinfo {author}
  {\bibfnamefont {C.~G.~R.}\ \bibnamefont {Geddes}}, \bibinfo {author}
  {\bibfnamefont {E.}~\bibnamefont {Esarey}}, \bibinfo {author} {\bibfnamefont
  {C.~B.}\ \bibnamefont {Schroeder}}, \ and\ \bibinfo {author} {\bibfnamefont
  {S.~M.}\ \bibnamefont {Hooker}},\ }\href {\doibase 10.1038/nphys418}
  {\bibfield  {journal} {\bibinfo  {journal} {Nature Physics}\ }\textbf
  {\bibinfo {volume} {2}},\ \bibinfo {pages} {696} (\bibinfo {year}
  {2006})}\BibitemShut {NoStop}%
\bibitem [{\citenamefont {Kneip}\ \emph {et~al.}(2009)\citenamefont {Kneip},
  \citenamefont {Nagel}, \citenamefont {Martins}, \citenamefont {Mangles},
  \citenamefont {Bellei}, \citenamefont {Chekhlov}, \citenamefont {Clarke},
  \citenamefont {Delerue}, \citenamefont {Divall}, \citenamefont {Doucas},
  \citenamefont {Ertel}, \citenamefont {Fiuza}, \citenamefont {Fonseca},
  \citenamefont {Foster}, \citenamefont {Hawkes}, \citenamefont {Hooker},
  \citenamefont {Krushelnick}, \citenamefont {Mori}, \citenamefont {Palmer},
  \citenamefont {Phuoc}, \citenamefont {Rajeev}, \citenamefont {Schreiber},
  \citenamefont {Streeter}, \citenamefont {Urner}, \citenamefont {Vieira},
  \citenamefont {Silva},\ and\ \citenamefont {Najmudin}}]{Kneip:2009}%
  \BibitemOpen
  \bibfield  {author} {\bibinfo {author} {\bibfnamefont {S.}~\bibnamefont
  {Kneip}}, \bibinfo {author} {\bibfnamefont {S.}~\bibnamefont {Nagel}},
  \bibinfo {author} {\bibfnamefont {S.}~\bibnamefont {Martins}}, \bibinfo
  {author} {\bibfnamefont {S.}~\bibnamefont {Mangles}}, \bibinfo {author}
  {\bibfnamefont {C.}~\bibnamefont {Bellei}}, \bibinfo {author} {\bibfnamefont
  {O.}~\bibnamefont {Chekhlov}}, \bibinfo {author} {\bibfnamefont
  {R.}~\bibnamefont {Clarke}}, \bibinfo {author} {\bibfnamefont
  {N.}~\bibnamefont {Delerue}}, \bibinfo {author} {\bibfnamefont
  {E.}~\bibnamefont {Divall}}, \bibinfo {author} {\bibfnamefont
  {G.}~\bibnamefont {Doucas}}, \bibinfo {author} {\bibfnamefont
  {K.}~\bibnamefont {Ertel}}, \bibinfo {author} {\bibfnamefont
  {F.}~\bibnamefont {Fiuza}}, \bibinfo {author} {\bibfnamefont
  {R.}~\bibnamefont {Fonseca}}, \bibinfo {author} {\bibfnamefont
  {P.}~\bibnamefont {Foster}}, \bibinfo {author} {\bibfnamefont
  {S.}~\bibnamefont {Hawkes}}, \bibinfo {author} {\bibfnamefont
  {C.}~\bibnamefont {Hooker}}, \bibinfo {author} {\bibfnamefont
  {K.}~\bibnamefont {Krushelnick}}, \bibinfo {author} {\bibfnamefont
  {W.}~\bibnamefont {Mori}}, \bibinfo {author} {\bibfnamefont {C.}~\bibnamefont
  {Palmer}}, \bibinfo {author} {\bibfnamefont {K.}~\bibnamefont {Phuoc}},
  \bibinfo {author} {\bibfnamefont {P.}~\bibnamefont {Rajeev}}, \bibinfo
  {author} {\bibfnamefont {J.}~\bibnamefont {Schreiber}}, \bibinfo {author}
  {\bibfnamefont {M.}~\bibnamefont {Streeter}}, \bibinfo {author}
  {\bibfnamefont {D.}~\bibnamefont {Urner}}, \bibinfo {author} {\bibfnamefont
  {J.}~\bibnamefont {Vieira}}, \bibinfo {author} {\bibfnamefont
  {L.}~\bibnamefont {Silva}}, \ and\ \bibinfo {author} {\bibfnamefont
  {Z.}~\bibnamefont {Najmudin}},\ }\href {\doibase
  10.1103/PhysRevLett.103.035002} {\bibfield  {journal} {\bibinfo  {journal}
  {Phys Rev Lett}\ }\textbf {\bibinfo {volume} {103}},\ \bibinfo {pages}
  {035002} (\bibinfo {year} {2009})}\BibitemShut {NoStop}%
\bibitem [{\citenamefont {Wang}\ \emph {et~al.}(2013)\citenamefont {Wang},
  \citenamefont {Zgadzaj}, \citenamefont {Fazel}, \citenamefont {Li},
  \citenamefont {Yi}, \citenamefont {Zhang}, \citenamefont {Henderson},
  \citenamefont {Chang}, \citenamefont {Korzekwa}, \citenamefont {Tsai},
  \citenamefont {Pai}, \citenamefont {Quevedo}, \citenamefont {Dyer},
  \citenamefont {Gaul}, \citenamefont {Martinez}, \citenamefont {Bernstein},
  \citenamefont {Borger}, \citenamefont {Spinks}, \citenamefont {Donovan},
  \citenamefont {Khudik}, \citenamefont {Shvets}, \citenamefont {Ditmire},\
  and\ \citenamefont {Downer}}]{Wang:2013el}%
  \BibitemOpen
  \bibfield  {author} {\bibinfo {author} {\bibfnamefont {X.}~\bibnamefont
  {Wang}}, \bibinfo {author} {\bibfnamefont {R.}~\bibnamefont {Zgadzaj}},
  \bibinfo {author} {\bibfnamefont {N.}~\bibnamefont {Fazel}}, \bibinfo
  {author} {\bibfnamefont {Z.}~\bibnamefont {Li}}, \bibinfo {author}
  {\bibfnamefont {S.~A.}\ \bibnamefont {Yi}}, \bibinfo {author} {\bibfnamefont
  {X.}~\bibnamefont {Zhang}}, \bibinfo {author} {\bibfnamefont
  {W.}~\bibnamefont {Henderson}}, \bibinfo {author} {\bibfnamefont {Y.~Y.}\
  \bibnamefont {Chang}}, \bibinfo {author} {\bibfnamefont {R.}~\bibnamefont
  {Korzekwa}}, \bibinfo {author} {\bibfnamefont {H.~E.}\ \bibnamefont {Tsai}},
  \bibinfo {author} {\bibfnamefont {C.~H.}\ \bibnamefont {Pai}}, \bibinfo
  {author} {\bibfnamefont {H.}~\bibnamefont {Quevedo}}, \bibinfo {author}
  {\bibfnamefont {G.}~\bibnamefont {Dyer}}, \bibinfo {author} {\bibfnamefont
  {E.}~\bibnamefont {Gaul}}, \bibinfo {author} {\bibfnamefont {M.}~\bibnamefont
  {Martinez}}, \bibinfo {author} {\bibfnamefont {A.~C.}\ \bibnamefont
  {Bernstein}}, \bibinfo {author} {\bibfnamefont {T.}~\bibnamefont {Borger}},
  \bibinfo {author} {\bibfnamefont {M.}~\bibnamefont {Spinks}}, \bibinfo
  {author} {\bibfnamefont {M.}~\bibnamefont {Donovan}}, \bibinfo {author}
  {\bibfnamefont {V.}~\bibnamefont {Khudik}}, \bibinfo {author} {\bibfnamefont
  {G.}~\bibnamefont {Shvets}}, \bibinfo {author} {\bibfnamefont
  {T.}~\bibnamefont {Ditmire}}, \ and\ \bibinfo {author} {\bibfnamefont
  {M.~C.}\ \bibnamefont {Downer}},\ }\href {\doibase 10.1038/ncomms2988}
  {\bibfield  {journal} {\bibinfo  {journal} {Nature Communications}\ }\textbf
  {\bibinfo {volume} {4}},\ \bibinfo {pages} {2988} (\bibinfo {year}
  {2013})}\BibitemShut {NoStop}%
\bibitem [{\citenamefont {Leemans}\ \emph {et~al.}(2014)\citenamefont
  {Leemans}, \citenamefont {Gonsalves}, \citenamefont {Mao}, \citenamefont
  {Nakamura}, \citenamefont {Benedetti}, \citenamefont {Schroeder},
  \citenamefont {Toth}, \citenamefont {Daniels}, \citenamefont {Mittelberger},
  \citenamefont {Bulanov}, \citenamefont {Vay}, \citenamefont {Geddes},\ and\
  \citenamefont {Esarey}}]{Leemans:2014kp}%
  \BibitemOpen
  \bibfield  {author} {\bibinfo {author} {\bibfnamefont {W.~P.}\ \bibnamefont
  {Leemans}}, \bibinfo {author} {\bibfnamefont {A.~J.}\ \bibnamefont
  {Gonsalves}}, \bibinfo {author} {\bibfnamefont {H.~S.}\ \bibnamefont {Mao}},
  \bibinfo {author} {\bibfnamefont {K.}~\bibnamefont {Nakamura}}, \bibinfo
  {author} {\bibfnamefont {C.}~\bibnamefont {Benedetti}}, \bibinfo {author}
  {\bibfnamefont {C.~B.}\ \bibnamefont {Schroeder}}, \bibinfo {author}
  {\bibfnamefont {C.}~\bibnamefont {Toth}}, \bibinfo {author} {\bibfnamefont
  {J.}~\bibnamefont {Daniels}}, \bibinfo {author} {\bibfnamefont {D.~E.}\
  \bibnamefont {Mittelberger}}, \bibinfo {author} {\bibfnamefont {S.~S.}\
  \bibnamefont {Bulanov}}, \bibinfo {author} {\bibfnamefont {J.~L.}\
  \bibnamefont {Vay}}, \bibinfo {author} {\bibfnamefont {C.~G.~R.}\
  \bibnamefont {Geddes}}, \ and\ \bibinfo {author} {\bibfnamefont
  {E.}~\bibnamefont {Esarey}},\ }\href {\doibase
  10.1103/PhysRevLett.113.245002} {\bibfield  {journal} {\bibinfo  {journal}
  {Phys Rev Lett}\ }\textbf {\bibinfo {volume} {113}},\ \bibinfo {pages}
  {245002} (\bibinfo {year} {2014})}\BibitemShut {NoStop}%
\bibitem [{\citenamefont {Buck}\ \emph {et~al.}(2011)\citenamefont {Buck},
  \citenamefont {Nicolai}, \citenamefont {Schmid}, \citenamefont {Sears},
  \citenamefont {S{\"a}vert}, \citenamefont {Mikhailova}, \citenamefont
  {Krausz}, \citenamefont {Kaluza},\ and\ \citenamefont {Veisz}}]{Buck:2011dg}%
  \BibitemOpen
  \bibfield  {author} {\bibinfo {author} {\bibfnamefont {A.}~\bibnamefont
  {Buck}}, \bibinfo {author} {\bibfnamefont {M.}~\bibnamefont {Nicolai}},
  \bibinfo {author} {\bibfnamefont {K.}~\bibnamefont {Schmid}}, \bibinfo
  {author} {\bibfnamefont {C.~M.~S.}\ \bibnamefont {Sears}}, \bibinfo {author}
  {\bibfnamefont {A.}~\bibnamefont {S{\"a}vert}}, \bibinfo {author}
  {\bibfnamefont {J.~M.}\ \bibnamefont {Mikhailova}}, \bibinfo {author}
  {\bibfnamefont {F.}~\bibnamefont {Krausz}}, \bibinfo {author} {\bibfnamefont
  {M.~C.}\ \bibnamefont {Kaluza}}, \ and\ \bibinfo {author} {\bibfnamefont
  {L.}~\bibnamefont {Veisz}},\ }\href {\doibase 10.1038/nphys1942} {\bibfield
  {journal} {\bibinfo  {journal} {Nat Phys}\ }\textbf {\bibinfo {volume} {7}},\
  \bibinfo {pages} {543} (\bibinfo {year} {2011})}\BibitemShut {NoStop}%
\bibitem [{\citenamefont {Lundh}\ \emph {et~al.}(2011)\citenamefont {Lundh},
  \citenamefont {Lim}, \citenamefont {Rechatin}, \citenamefont {Ammoura},
  \citenamefont {Ben-Ismail}, \citenamefont {Davoine}, \citenamefont {Gallot},
  \citenamefont {Goddet}, \citenamefont {Lefebvre}, \citenamefont {Malka},\
  and\ \citenamefont {Faure}}]{Lundh:2011b}%
  \BibitemOpen
  \bibfield  {author} {\bibinfo {author} {\bibfnamefont {O.}~\bibnamefont
  {Lundh}}, \bibinfo {author} {\bibfnamefont {J.}~\bibnamefont {Lim}}, \bibinfo
  {author} {\bibfnamefont {C.}~\bibnamefont {Rechatin}}, \bibinfo {author}
  {\bibfnamefont {L.}~\bibnamefont {Ammoura}}, \bibinfo {author} {\bibfnamefont
  {A.}~\bibnamefont {Ben-Ismail}}, \bibinfo {author} {\bibfnamefont
  {X.}~\bibnamefont {Davoine}}, \bibinfo {author} {\bibfnamefont
  {G.}~\bibnamefont {Gallot}}, \bibinfo {author} {\bibfnamefont {J.-P.}\
  \bibnamefont {Goddet}}, \bibinfo {author} {\bibfnamefont {E.}~\bibnamefont
  {Lefebvre}}, \bibinfo {author} {\bibfnamefont {V.}~\bibnamefont {Malka}}, \
  and\ \bibinfo {author} {\bibfnamefont {J.}~\bibnamefont {Faure}},\ }\href
  {\doibase 10.1038/nphys1872} {\bibfield  {journal} {\bibinfo  {journal} {Nat.
  Phys.}\ }\textbf {\bibinfo {volume} {7}},\ \bibinfo {pages} {219} (\bibinfo
  {year} {2011})}\BibitemShut {NoStop}%
\bibitem [{\citenamefont {Heigoldt}\ \emph {et~al.}(2015)\citenamefont
  {Heigoldt}, \citenamefont {Popp}, \citenamefont {Khrennikov}, \citenamefont
  {Wenz}, \citenamefont {Chou}, \citenamefont {Karsch}, \citenamefont
  {Bajlekov}, \citenamefont {Hooker},\ and\ \citenamefont
  {Schmidt}}]{Heigoldt:2015cd}%
  \BibitemOpen
  \bibfield  {author} {\bibinfo {author} {\bibfnamefont {M.}~\bibnamefont
  {Heigoldt}}, \bibinfo {author} {\bibfnamefont {A.}~\bibnamefont {Popp}},
  \bibinfo {author} {\bibfnamefont {K.}~\bibnamefont {Khrennikov}}, \bibinfo
  {author} {\bibfnamefont {J.}~\bibnamefont {Wenz}}, \bibinfo {author}
  {\bibfnamefont {S.-W.}\ \bibnamefont {Chou}}, \bibinfo {author}
  {\bibfnamefont {S.}~\bibnamefont {Karsch}}, \bibinfo {author} {\bibfnamefont
  {S.~I.}\ \bibnamefont {Bajlekov}}, \bibinfo {author} {\bibfnamefont {S.~M.}\
  \bibnamefont {Hooker}}, \ and\ \bibinfo {author} {\bibfnamefont
  {B.}~\bibnamefont {Schmidt}},\ }\href {\doibase
  10.1103/PhysRevSTAB.18.121302} {\bibfield  {journal} {\bibinfo  {journal}
  {Phys Rev Spec Top-Ac}\ }\textbf {\bibinfo {volume} {18}},\ \bibinfo {pages}
  {121302} (\bibinfo {year} {2015})}\BibitemShut {NoStop}%
\bibitem [{\citenamefont {Schlenvoigt}\ \emph {et~al.}(2008)\citenamefont
  {Schlenvoigt}, \citenamefont {Haupt}, \citenamefont {Debus}, \citenamefont
  {Budde}, \citenamefont {Jackel}, \citenamefont {Pfotenhauer}, \citenamefont
  {Schwoerer}, \citenamefont {Rohwer}, \citenamefont {Gallacher}, \citenamefont
  {Brunetti}, \citenamefont {Shanks}, \citenamefont {Wiggins},\ and\
  \citenamefont {Jaroszynski}}]{Schlenvoigt:2008}%
  \BibitemOpen
  \bibfield  {author} {\bibinfo {author} {\bibfnamefont {H.-P.}\ \bibnamefont
  {Schlenvoigt}}, \bibinfo {author} {\bibfnamefont {K.}~\bibnamefont {Haupt}},
  \bibinfo {author} {\bibfnamefont {A.}~\bibnamefont {Debus}}, \bibinfo
  {author} {\bibfnamefont {F.}~\bibnamefont {Budde}}, \bibinfo {author}
  {\bibfnamefont {O.}~\bibnamefont {Jackel}}, \bibinfo {author} {\bibfnamefont
  {S.}~\bibnamefont {Pfotenhauer}}, \bibinfo {author} {\bibfnamefont
  {H.}~\bibnamefont {Schwoerer}}, \bibinfo {author} {\bibfnamefont
  {E.}~\bibnamefont {Rohwer}}, \bibinfo {author} {\bibfnamefont {J.~G.}\
  \bibnamefont {Gallacher}}, \bibinfo {author} {\bibfnamefont {E.}~\bibnamefont
  {Brunetti}}, \bibinfo {author} {\bibfnamefont {R.~P.}\ \bibnamefont
  {Shanks}}, \bibinfo {author} {\bibfnamefont {S.~M.}\ \bibnamefont {Wiggins}},
  \ and\ \bibinfo {author} {\bibfnamefont {D.~A.}\ \bibnamefont
  {Jaroszynski}},\ }\href {<Go to ISI>://000253724500017} {\bibfield  {journal}
  {\bibinfo  {journal} {Nature Physics}\ }\textbf {\bibinfo {volume} {4}},\
  \bibinfo {pages} {130} (\bibinfo {year} {2008})}\BibitemShut {NoStop}%
\bibitem [{\citenamefont {Fuchs}\ \emph {et~al.}(2009)\citenamefont {Fuchs},
  \citenamefont {Weingartner}, \citenamefont {Popp}, \citenamefont {Major},
  \citenamefont {Becker}, \citenamefont {Osterhoff}, \citenamefont {Cortrie},
  \citenamefont {Zeitler}, \citenamefont {Hoerlein}, \citenamefont {Tsakiris},
  \citenamefont {Schramm}, \citenamefont {Rowlands-Rees}, \citenamefont
  {Hooker}, \citenamefont {Habs}, \citenamefont {Krausz}, \citenamefont
  {Karsch},\ and\ \citenamefont {Gruener}}]{Fuchs:2009}%
  \BibitemOpen
  \bibfield  {author} {\bibinfo {author} {\bibfnamefont {M.}~\bibnamefont
  {Fuchs}}, \bibinfo {author} {\bibfnamefont {R.}~\bibnamefont {Weingartner}},
  \bibinfo {author} {\bibfnamefont {A.}~\bibnamefont {Popp}}, \bibinfo {author}
  {\bibfnamefont {Z.}~\bibnamefont {Major}}, \bibinfo {author} {\bibfnamefont
  {S.}~\bibnamefont {Becker}}, \bibinfo {author} {\bibfnamefont
  {J.}~\bibnamefont {Osterhoff}}, \bibinfo {author} {\bibfnamefont
  {I.}~\bibnamefont {Cortrie}}, \bibinfo {author} {\bibfnamefont
  {B.}~\bibnamefont {Zeitler}}, \bibinfo {author} {\bibfnamefont
  {R.}~\bibnamefont {Hoerlein}}, \bibinfo {author} {\bibfnamefont {G.~D.}\
  \bibnamefont {Tsakiris}}, \bibinfo {author} {\bibfnamefont {U.}~\bibnamefont
  {Schramm}}, \bibinfo {author} {\bibfnamefont {T.~P.}\ \bibnamefont
  {Rowlands-Rees}}, \bibinfo {author} {\bibfnamefont {S.~M.}\ \bibnamefont
  {Hooker}}, \bibinfo {author} {\bibfnamefont {D.}~\bibnamefont {Habs}},
  \bibinfo {author} {\bibfnamefont {F.}~\bibnamefont {Krausz}}, \bibinfo
  {author} {\bibfnamefont {S.}~\bibnamefont {Karsch}}, \ and\ \bibinfo {author}
  {\bibfnamefont {F.}~\bibnamefont {Gruener}},\ }\href {\doibase
  10.1038/NPHYS1404} {\bibfield  {journal} {\bibinfo  {journal} {Nature
  Physics}\ }\textbf {\bibinfo {volume} {5}},\ \bibinfo {pages} {826} (\bibinfo
  {year} {2009})}\BibitemShut {NoStop}%
\bibitem [{\citenamefont {Kneip}\ \emph {et~al.}(2010)\citenamefont {Kneip},
  \citenamefont {Mcguffey}, \citenamefont {Martins}, \citenamefont {Martins},
  \citenamefont {Bellei}, \citenamefont {Chvykov}, \citenamefont {Dollar},
  \citenamefont {Fonseca}, \citenamefont {Huntington}, \citenamefont
  {Kalintchenko}, \citenamefont {Maksimchuk}, \citenamefont {Mangles},
  \citenamefont {Matsuoka}, \citenamefont {Nagel}, \citenamefont {Palmer},
  \citenamefont {Schreiber}, \citenamefont {Phuoc}, \citenamefont {Thomas},
  \citenamefont {Yanovsky}, \citenamefont {Silva}, \citenamefont
  {Krushelnick},\ and\ \citenamefont {Najmudin}}]{Kneip:2010}%
  \BibitemOpen
  \bibfield  {author} {\bibinfo {author} {\bibfnamefont {S.}~\bibnamefont
  {Kneip}}, \bibinfo {author} {\bibfnamefont {C.}~\bibnamefont {Mcguffey}},
  \bibinfo {author} {\bibfnamefont {J.~L.}\ \bibnamefont {Martins}}, \bibinfo
  {author} {\bibfnamefont {S.~F.}\ \bibnamefont {Martins}}, \bibinfo {author}
  {\bibfnamefont {C.}~\bibnamefont {Bellei}}, \bibinfo {author} {\bibfnamefont
  {V.}~\bibnamefont {Chvykov}}, \bibinfo {author} {\bibfnamefont
  {F.}~\bibnamefont {Dollar}}, \bibinfo {author} {\bibfnamefont
  {R.}~\bibnamefont {Fonseca}}, \bibinfo {author} {\bibfnamefont
  {C.}~\bibnamefont {Huntington}}, \bibinfo {author} {\bibfnamefont
  {G.}~\bibnamefont {Kalintchenko}}, \bibinfo {author} {\bibfnamefont
  {A.}~\bibnamefont {Maksimchuk}}, \bibinfo {author} {\bibfnamefont {S.~P.~D.}\
  \bibnamefont {Mangles}}, \bibinfo {author} {\bibfnamefont {T.}~\bibnamefont
  {Matsuoka}}, \bibinfo {author} {\bibfnamefont {S.~R.}\ \bibnamefont {Nagel}},
  \bibinfo {author} {\bibfnamefont {C.~A.~J.}\ \bibnamefont {Palmer}}, \bibinfo
  {author} {\bibfnamefont {J.}~\bibnamefont {Schreiber}}, \bibinfo {author}
  {\bibfnamefont {K.~T.}\ \bibnamefont {Phuoc}}, \bibinfo {author}
  {\bibfnamefont {A.~G.~R.}\ \bibnamefont {Thomas}}, \bibinfo {author}
  {\bibfnamefont {V.}~\bibnamefont {Yanovsky}}, \bibinfo {author}
  {\bibfnamefont {L.~O.}\ \bibnamefont {Silva}}, \bibinfo {author}
  {\bibfnamefont {K.}~\bibnamefont {Krushelnick}}, \ and\ \bibinfo {author}
  {\bibfnamefont {Z.}~\bibnamefont {Najmudin}},\ }\href {\doibase
  10.1038/nphys1789} {\bibfield  {journal} {\bibinfo  {journal} {Nat. Phys.}\
  }\textbf {\bibinfo {volume} {6}},\ \bibinfo {pages} {980} (\bibinfo {year}
  {2010})}\BibitemShut {NoStop}%
\bibitem [{\citenamefont {Cipiccia}\ \emph {et~al.}(2011)\citenamefont
  {Cipiccia}, \citenamefont {Islam}, \citenamefont {Ersfeld}, \citenamefont
  {Shanks}, \citenamefont {Brunetti}, \citenamefont {Vieux}, \citenamefont
  {Yang}, \citenamefont {Issac}, \citenamefont {Wiggins}, \citenamefont
  {Welsh}, \citenamefont {Anania}, \citenamefont {Maneuski}, \citenamefont
  {Montgomery}, \citenamefont {Smith}, \citenamefont {Hoek}, \citenamefont
  {Hamilton}, \citenamefont {Lemos}, \citenamefont {Symes}, \citenamefont
  {Rajeev}, \citenamefont {Shea}, \citenamefont {Dias},\ and\ \citenamefont
  {Jaroszynski}}]{Cipiccia:2011}%
  \BibitemOpen
  \bibfield  {author} {\bibinfo {author} {\bibfnamefont {S.}~\bibnamefont
  {Cipiccia}}, \bibinfo {author} {\bibfnamefont {M.~R.}\ \bibnamefont {Islam}},
  \bibinfo {author} {\bibfnamefont {B.}~\bibnamefont {Ersfeld}}, \bibinfo
  {author} {\bibfnamefont {R.~P.}\ \bibnamefont {Shanks}}, \bibinfo {author}
  {\bibfnamefont {E.}~\bibnamefont {Brunetti}}, \bibinfo {author}
  {\bibfnamefont {G.}~\bibnamefont {Vieux}}, \bibinfo {author} {\bibfnamefont
  {X.}~\bibnamefont {Yang}}, \bibinfo {author} {\bibfnamefont {R.~C.}\
  \bibnamefont {Issac}}, \bibinfo {author} {\bibfnamefont {S.~M.}\ \bibnamefont
  {Wiggins}}, \bibinfo {author} {\bibfnamefont {G.~H.}\ \bibnamefont {Welsh}},
  \bibinfo {author} {\bibfnamefont {M.-P.}\ \bibnamefont {Anania}}, \bibinfo
  {author} {\bibfnamefont {D.}~\bibnamefont {Maneuski}}, \bibinfo {author}
  {\bibfnamefont {R.}~\bibnamefont {Montgomery}}, \bibinfo {author}
  {\bibfnamefont {G.}~\bibnamefont {Smith}}, \bibinfo {author} {\bibfnamefont
  {M.}~\bibnamefont {Hoek}}, \bibinfo {author} {\bibfnamefont {D.~J.}\
  \bibnamefont {Hamilton}}, \bibinfo {author} {\bibfnamefont {N.~R.~C.}\
  \bibnamefont {Lemos}}, \bibinfo {author} {\bibfnamefont {D.}~\bibnamefont
  {Symes}}, \bibinfo {author} {\bibfnamefont {P.~P.}\ \bibnamefont {Rajeev}},
  \bibinfo {author} {\bibfnamefont {V.~O.}\ \bibnamefont {Shea}}, \bibinfo
  {author} {\bibfnamefont {J.~M.}\ \bibnamefont {Dias}}, \ and\ \bibinfo
  {author} {\bibfnamefont {D.~A.}\ \bibnamefont {Jaroszynski}},\ }\href
  {\doibase 10.1038/nphys2090} {\bibfield  {journal} {\bibinfo  {journal}
  {Nature Physics}\ }\textbf {\bibinfo {volume} {7}},\ \bibinfo {pages} {867}
  (\bibinfo {year} {2011})}\BibitemShut {NoStop}%
\bibitem [{\citenamefont {Phuoc}\ \emph {et~al.}(2012)\citenamefont {Phuoc},
  \citenamefont {Corde}, \citenamefont {Thaury}, \citenamefont {Malka},
  \citenamefont {Tafzi}, \citenamefont {Goddet}, \citenamefont {Shah},
  \citenamefont {Sebban},\ and\ \citenamefont {Rousse}}]{Phuoc:2012vb}%
  \BibitemOpen
  \bibfield  {author} {\bibinfo {author} {\bibfnamefont {K.~T.}\ \bibnamefont
  {Phuoc}}, \bibinfo {author} {\bibfnamefont {S.}~\bibnamefont {Corde}},
  \bibinfo {author} {\bibfnamefont {C.}~\bibnamefont {Thaury}}, \bibinfo
  {author} {\bibfnamefont {V.}~\bibnamefont {Malka}}, \bibinfo {author}
  {\bibfnamefont {A.}~\bibnamefont {Tafzi}}, \bibinfo {author} {\bibfnamefont
  {J.-P.}\ \bibnamefont {Goddet}}, \bibinfo {author} {\bibfnamefont {R.~C.}\
  \bibnamefont {Shah}}, \bibinfo {author} {\bibfnamefont {S.}~\bibnamefont
  {Sebban}}, \ and\ \bibinfo {author} {\bibfnamefont {A.}~\bibnamefont
  {Rousse}},\ }\href
  {http://www.nature.com/nphoton/journal/v6/n5/abs/nphoton.2012.82.html}
  {\bibfield  {journal} {\bibinfo  {journal} {Nature Photonics}\ }\textbf
  {\bibinfo {volume} {6}},\ \bibinfo {pages} {308} (\bibinfo {year}
  {2012})}\BibitemShut {NoStop}%
\bibitem [{\citenamefont {Powers}\ \emph {et~al.}(2013)\citenamefont {Powers},
  \citenamefont {Ghebregziabher}, \citenamefont {Golovin}, \citenamefont {Liu},
  \citenamefont {Chen}, \citenamefont {Banerjee}, \citenamefont {Zhang},\ and\
  \citenamefont {Umstadter}}]{Powers:2013bx}%
  \BibitemOpen
  \bibfield  {author} {\bibinfo {author} {\bibfnamefont {N.~D.}\ \bibnamefont
  {Powers}}, \bibinfo {author} {\bibfnamefont {I.}~\bibnamefont
  {Ghebregziabher}}, \bibinfo {author} {\bibfnamefont {G.}~\bibnamefont
  {Golovin}}, \bibinfo {author} {\bibfnamefont {C.}~\bibnamefont {Liu}},
  \bibinfo {author} {\bibfnamefont {S.}~\bibnamefont {Chen}}, \bibinfo {author}
  {\bibfnamefont {S.}~\bibnamefont {Banerjee}}, \bibinfo {author}
  {\bibfnamefont {J.}~\bibnamefont {Zhang}}, \ and\ \bibinfo {author}
  {\bibfnamefont {D.~P.}\ \bibnamefont {Umstadter}},\ }\href {\doibase
  10.1038/nphoton.2013.314} {\bibfield  {journal} {\bibinfo  {journal} {Nature
  Photonics}\ }\textbf {\bibinfo {volume} {8}},\ \bibinfo {pages} {28}
  (\bibinfo {year} {2013})}\BibitemShut {NoStop}%
\bibitem [{\citenamefont {Khrennikov}\ \emph {et~al.}(2015)\citenamefont
  {Khrennikov}, \citenamefont {Wenz}, \citenamefont {Buck}, \citenamefont {Xu},
  \citenamefont {Heigoldt}, \citenamefont {Veisz},\ and\ \citenamefont
  {Karsch}}]{Khrennikov:2015gx}%
  \BibitemOpen
  \bibfield  {author} {\bibinfo {author} {\bibfnamefont {K.}~\bibnamefont
  {Khrennikov}}, \bibinfo {author} {\bibfnamefont {J.}~\bibnamefont {Wenz}},
  \bibinfo {author} {\bibfnamefont {A.}~\bibnamefont {Buck}}, \bibinfo {author}
  {\bibfnamefont {J.}~\bibnamefont {Xu}}, \bibinfo {author} {\bibfnamefont
  {M.}~\bibnamefont {Heigoldt}}, \bibinfo {author} {\bibfnamefont
  {L.}~\bibnamefont {Veisz}}, \ and\ \bibinfo {author} {\bibfnamefont
  {S.}~\bibnamefont {Karsch}},\ }\href {\doibase
  10.1103/PhysRevLett.114.195003} {\bibfield  {journal} {\bibinfo  {journal}
  {Phys. Rev. Lett.}\ }\textbf {\bibinfo {volume} {114}},\ \bibinfo {pages}
  {195003} (\bibinfo {year} {2015})}\BibitemShut {NoStop}%
\bibitem [{\citenamefont {Esarey}\ \emph {et~al.}(2009)\citenamefont {Esarey},
  \citenamefont {Schroeder},\ and\ \citenamefont {Leemans}}]{Esarey:2009}%
  \BibitemOpen
  \bibfield  {author} {\bibinfo {author} {\bibfnamefont {E.}~\bibnamefont
  {Esarey}}, \bibinfo {author} {\bibfnamefont {C.~B.}\ \bibnamefont
  {Schroeder}}, \ and\ \bibinfo {author} {\bibfnamefont {W.~P.}\ \bibnamefont
  {Leemans}},\ }\href {\doibase 10.1103/RevModPhys.81.1229} {\bibfield
  {journal} {\bibinfo  {journal} {Rev. Mod. Phys.}\ }\textbf {\bibinfo {volume}
  {81}},\ \bibinfo {pages} {1229} (\bibinfo {year} {2009})}\BibitemShut
  {NoStop}%
\bibitem [{\citenamefont {Cros}\ \emph {et~al.}(2017)\citenamefont {Cros},
  \citenamefont {Audet}, \citenamefont {Hooker}, \citenamefont {Koester},
  \citenamefont {Tomassini}, \citenamefont {Labate}, \citenamefont {Poder},\
  and\ \citenamefont {Najmudin}}]{Cros:2017}%
  \BibitemOpen
  \bibfield  {author} {\bibinfo {author} {\bibfnamefont {B.}~\bibnamefont
  {Cros}}, \bibinfo {author} {\bibfnamefont {T.}~\bibnamefont {Audet}},
  \bibinfo {author} {\bibfnamefont {S.~M.}\ \bibnamefont {Hooker}}, \bibinfo
  {author} {\bibfnamefont {P.}~\bibnamefont {Koester}}, \bibinfo {author}
  {\bibfnamefont {P.}~\bibnamefont {Tomassini}}, \bibinfo {author}
  {\bibfnamefont {L.}~\bibnamefont {Labate}}, \bibinfo {author} {\bibfnamefont
  {K.}~\bibnamefont {Poder}}, \ and\ \bibinfo {author} {\bibfnamefont
  {Z.}~\bibnamefont {Najmudin}},\ }\href@noop {} {\enquote {\bibinfo {title}
  {Design for an electron injector and a laser plasma stage proposed {EuPRAXIA}
  {M}ilestone {R}eport 3.1},}\ } (\bibinfo {year} {2017})\BibitemShut {NoStop}%
\bibitem [{\citenamefont {Leemans}\ \emph {et~al.}(2011)\citenamefont
  {Leemans}, \citenamefont {Duarte}, \citenamefont {Esarey}, \citenamefont
  {Fournier}, \citenamefont {Geddes}, \citenamefont {Lockhart}, \citenamefont
  {Schroeder}, \citenamefont {Toth}, \citenamefont {Vay},\ and\ \citenamefont
  {Zimmermann}}]{Leemans:2011um}%
  \BibitemOpen
  \bibfield  {author} {\bibinfo {author} {\bibfnamefont {W.~P.}\ \bibnamefont
  {Leemans}}, \bibinfo {author} {\bibfnamefont {R.}~\bibnamefont {Duarte}},
  \bibinfo {author} {\bibfnamefont {E.}~\bibnamefont {Esarey}}, \bibinfo
  {author} {\bibfnamefont {S.}~\bibnamefont {Fournier}}, \bibinfo {author}
  {\bibfnamefont {C.}~\bibnamefont {Geddes}}, \bibinfo {author} {\bibfnamefont
  {D.}~\bibnamefont {Lockhart}}, \bibinfo {author} {\bibfnamefont {C.~B.}\
  \bibnamefont {Schroeder}}, \bibinfo {author} {\bibfnamefont {C.}~\bibnamefont
  {Toth}}, \bibinfo {author} {\bibfnamefont {J.~L.}\ \bibnamefont {Vay}}, \
  and\ \bibinfo {author} {\bibfnamefont {S.}~\bibnamefont {Zimmermann}},\ }in\
  \href@noop {} {\emph {\bibinfo {booktitle} {Particle Accelerator
  Conference}}}\ (\bibinfo {year} {2011})\ pp.\ \bibinfo {pages}
  {1416--1420}\BibitemShut {NoStop}%
\bibitem [{\citenamefont {Cros}\ \emph {et~al.}(2002)\citenamefont {Cros},
  \citenamefont {Courtois}, \citenamefont {Matthieussent}, \citenamefont
  {Di~Bernardo}, \citenamefont {Batani}, \citenamefont {Andreev},\ and\
  \citenamefont {Kuznetsov}}]{Cros:2002}%
  \BibitemOpen
  \bibfield  {author} {\bibinfo {author} {\bibfnamefont {B.}~\bibnamefont
  {Cros}}, \bibinfo {author} {\bibfnamefont {C.}~\bibnamefont {Courtois}},
  \bibinfo {author} {\bibfnamefont {G.}~\bibnamefont {Matthieussent}}, \bibinfo
  {author} {\bibfnamefont {A.}~\bibnamefont {Di~Bernardo}}, \bibinfo {author}
  {\bibfnamefont {D.}~\bibnamefont {Batani}}, \bibinfo {author} {\bibfnamefont
  {N.}~\bibnamefont {Andreev}}, \ and\ \bibinfo {author} {\bibfnamefont
  {S.}~\bibnamefont {Kuznetsov}},\ }\href {\doibase 10.1103/PhysRevE.65.026405}
  {\bibfield  {journal} {\bibinfo  {journal} {Phys Rev E}\ }\textbf {\bibinfo
  {volume} {65}},\ \bibinfo {pages} {026405} (\bibinfo {year}
  {2002})}\BibitemShut {NoStop}%
\bibitem [{\citenamefont {Dorchies}\ \emph {et~al.}(1999)\citenamefont
  {Dorchies}, \citenamefont {Marques}, \citenamefont {Cros}, \citenamefont
  {Matthieussent}, \citenamefont {Courtois}, \citenamefont {Velikoroussov},
  \citenamefont {Audebert}, \citenamefont {Geindre}, \citenamefont {Rebibo},
  \citenamefont {Hamoniaux},\ and\ \citenamefont
  {Amiranoff}}]{Dorchies:1999vb}%
  \BibitemOpen
  \bibfield  {author} {\bibinfo {author} {\bibfnamefont {F.}~\bibnamefont
  {Dorchies}}, \bibinfo {author} {\bibfnamefont {J.~R.}\ \bibnamefont
  {Marques}}, \bibinfo {author} {\bibfnamefont {B.}~\bibnamefont {Cros}},
  \bibinfo {author} {\bibfnamefont {G.}~\bibnamefont {Matthieussent}}, \bibinfo
  {author} {\bibfnamefont {C.}~\bibnamefont {Courtois}}, \bibinfo {author}
  {\bibfnamefont {T.}~\bibnamefont {Velikoroussov}}, \bibinfo {author}
  {\bibfnamefont {P.}~\bibnamefont {Audebert}}, \bibinfo {author}
  {\bibfnamefont {J.~P.}\ \bibnamefont {Geindre}}, \bibinfo {author}
  {\bibfnamefont {S.}~\bibnamefont {Rebibo}}, \bibinfo {author} {\bibfnamefont
  {G.}~\bibnamefont {Hamoniaux}}, \ and\ \bibinfo {author} {\bibfnamefont
  {F.}~\bibnamefont {Amiranoff}},\ }\href@noop {} {\bibfield  {journal}
  {\bibinfo  {journal} {Phys Rev Lett}\ }\textbf {\bibinfo {volume} {82}},\
  \bibinfo {pages} {4655} (\bibinfo {year} {1999})}\BibitemShut {NoStop}%
\bibitem [{\citenamefont {Hansson}\ \emph {et~al.}(2014)\citenamefont
  {Hansson}, \citenamefont {Senje}, \citenamefont {Persson}, \citenamefont
  {Lundh}, \citenamefont {Wahlstr{\"o}m}, \citenamefont {Desforges},
  \citenamefont {Ju}, \citenamefont {Audet}, \citenamefont {Cros},
  \citenamefont {Dobosz~Dufr{\'e}noy},\ and\ \citenamefont
  {Monot}}]{Hansson:2014fk}%
  \BibitemOpen
  \bibfield  {author} {\bibinfo {author} {\bibfnamefont {M.}~\bibnamefont
  {Hansson}}, \bibinfo {author} {\bibfnamefont {L.}~\bibnamefont {Senje}},
  \bibinfo {author} {\bibfnamefont {A.}~\bibnamefont {Persson}}, \bibinfo
  {author} {\bibfnamefont {O.}~\bibnamefont {Lundh}}, \bibinfo {author}
  {\bibfnamefont {C.-G.}\ \bibnamefont {Wahlstr{\"o}m}}, \bibinfo {author}
  {\bibfnamefont {F.~G.}\ \bibnamefont {Desforges}}, \bibinfo {author}
  {\bibfnamefont {J.}~\bibnamefont {Ju}}, \bibinfo {author} {\bibfnamefont
  {T.~L.}\ \bibnamefont {Audet}}, \bibinfo {author} {\bibfnamefont
  {B.}~\bibnamefont {Cros}}, \bibinfo {author} {\bibfnamefont {S.}~\bibnamefont
  {Dobosz~Dufr{\'e}noy}}, \ and\ \bibinfo {author} {\bibfnamefont
  {P.}~\bibnamefont {Monot}},\ }\href {\doibase 10.1103/PhysRevSTAB.17.031303}
  {\bibfield  {journal} {\bibinfo  {journal} {Phys Rev Spec Top-Ac}\ }\textbf
  {\bibinfo {volume} {17}},\ \bibinfo {pages} {031303} (\bibinfo {year}
  {2014})}\BibitemShut {NoStop}%
\bibitem [{\citenamefont {Ehrlich}\ \emph {et~al.}(1996)\citenamefont
  {Ehrlich}, \citenamefont {Cohen}, \citenamefont {Zigler}, \citenamefont
  {Krall}, \citenamefont {Sprangle},\ and\ \citenamefont
  {Esarey}}]{Ehrlich:1996}%
  \BibitemOpen
  \bibfield  {author} {\bibinfo {author} {\bibfnamefont {Y.}~\bibnamefont
  {Ehrlich}}, \bibinfo {author} {\bibfnamefont {C.}~\bibnamefont {Cohen}},
  \bibinfo {author} {\bibfnamefont {A.}~\bibnamefont {Zigler}}, \bibinfo
  {author} {\bibfnamefont {J.}~\bibnamefont {Krall}}, \bibinfo {author}
  {\bibfnamefont {P.}~\bibnamefont {Sprangle}}, \ and\ \bibinfo {author}
  {\bibfnamefont {E.}~\bibnamefont {Esarey}},\ }\href {\doibase
  10.1103/PhysRevLett.77.4186} {\bibfield  {journal} {\bibinfo  {journal} {Phys
  Rev Lett}\ }\textbf {\bibinfo {volume} {77}},\ \bibinfo {pages} {4186}
  (\bibinfo {year} {1996})}\BibitemShut {NoStop}%
\bibitem [{\citenamefont {Hosokai}\ \emph {et~al.}(2000)\citenamefont
  {Hosokai}, \citenamefont {Kando}, \citenamefont {Dewa}, \citenamefont
  {Kotaki}, \citenamefont {Kondo}, \citenamefont {Hasegawa}, \citenamefont
  {Nakajima},\ and\ \citenamefont {Horioka}}]{Hosokai:2000}%
  \BibitemOpen
  \bibfield  {author} {\bibinfo {author} {\bibfnamefont {T.}~\bibnamefont
  {Hosokai}}, \bibinfo {author} {\bibfnamefont {M.}~\bibnamefont {Kando}},
  \bibinfo {author} {\bibfnamefont {H.}~\bibnamefont {Dewa}}, \bibinfo {author}
  {\bibfnamefont {H.}~\bibnamefont {Kotaki}}, \bibinfo {author} {\bibfnamefont
  {S.}~\bibnamefont {Kondo}}, \bibinfo {author} {\bibfnamefont
  {N.}~\bibnamefont {Hasegawa}}, \bibinfo {author} {\bibfnamefont
  {K.}~\bibnamefont {Nakajima}}, \ and\ \bibinfo {author} {\bibfnamefont
  {K.}~\bibnamefont {Horioka}},\ }\href@noop {} {\bibfield  {journal} {\bibinfo
   {journal} {Opt Lett}\ }\textbf {\bibinfo {volume} {25}},\ \bibinfo {pages}
  {10} (\bibinfo {year} {2000})}\BibitemShut {NoStop}%
\bibitem [{\citenamefont {Lopes}\ \emph {et~al.}(2003)\citenamefont {Lopes},
  \citenamefont {Figueira}, \citenamefont {Silva}, \citenamefont {Dias},
  \citenamefont {Fonseca}, \citenamefont {Cardoso}, \citenamefont {Russo},
  \citenamefont {Carias}, \citenamefont {Mendes}, \citenamefont {Vieira},\ and\
  \citenamefont {Mendonca}}]{Lopes:2003}%
  \BibitemOpen
  \bibfield  {author} {\bibinfo {author} {\bibfnamefont {N.~C.}\ \bibnamefont
  {Lopes}}, \bibinfo {author} {\bibfnamefont {G.}~\bibnamefont {Figueira}},
  \bibinfo {author} {\bibfnamefont {L.~O.}\ \bibnamefont {Silva}}, \bibinfo
  {author} {\bibfnamefont {J.~M.}\ \bibnamefont {Dias}}, \bibinfo {author}
  {\bibfnamefont {R.}~\bibnamefont {Fonseca}}, \bibinfo {author} {\bibfnamefont
  {L.}~\bibnamefont {Cardoso}}, \bibinfo {author} {\bibfnamefont
  {C.}~\bibnamefont {Russo}}, \bibinfo {author} {\bibfnamefont
  {C.}~\bibnamefont {Carias}}, \bibinfo {author} {\bibfnamefont
  {G.}~\bibnamefont {Mendes}}, \bibinfo {author} {\bibfnamefont
  {J.}~\bibnamefont {Vieira}}, \ and\ \bibinfo {author} {\bibfnamefont {J.~T.}\
  \bibnamefont {Mendonca}},\ }\href {\doibase 10.1103/PhysRevE.68.035402}
  {\bibfield  {journal} {\bibinfo  {journal} {Phys Rev E}\ }\textbf {\bibinfo
  {volume} {68}},\ \bibinfo {pages} {035402} (\bibinfo {year}
  {2003})}\BibitemShut {NoStop}%
\bibitem [{\citenamefont {Durfee}\ and\ \citenamefont
  {Milchberg}(1993)}]{Durfee:1993}%
  \BibitemOpen
  \bibfield  {author} {\bibinfo {author} {\bibfnamefont {C.~G.}\ \bibnamefont
  {Durfee}}\ and\ \bibinfo {author} {\bibfnamefont {H.~M.}\ \bibnamefont
  {Milchberg}},\ }\href {\doibase 10.1103/PhysRevLett.71.2409} {\bibfield
  {journal} {\bibinfo  {journal} {Phys Rev Lett}\ }\textbf {\bibinfo {volume}
  {71}},\ \bibinfo {pages} {2409} (\bibinfo {year} {1993})}\BibitemShut
  {NoStop}%
\bibitem [{\citenamefont {Durfee}\ \emph {et~al.}(1994)\citenamefont {Durfee},
  \citenamefont {Lynch},\ and\ \citenamefont {Milchberg}}]{Durfee:1994wz}%
  \BibitemOpen
  \bibfield  {author} {\bibinfo {author} {\bibfnamefont {C.~G.}\ \bibnamefont
  {Durfee}}, \bibinfo {author} {\bibfnamefont {J.}~\bibnamefont {Lynch}}, \
  and\ \bibinfo {author} {\bibfnamefont {H.~M.}\ \bibnamefont {Milchberg}},\
  }\href@noop {} {\bibfield  {journal} {\bibinfo  {journal} {Opt Lett}\
  }\textbf {\bibinfo {volume} {19}},\ \bibinfo {pages} {1937} (\bibinfo {year}
  {1994})}\BibitemShut {NoStop}%
\bibitem [{\citenamefont {Durfee}\ \emph {et~al.}(1995)\citenamefont {Durfee},
  \citenamefont {Lynch},\ and\ \citenamefont {Milchberg}}]{Durfee:1995gr}%
  \BibitemOpen
  \bibfield  {author} {\bibinfo {author} {\bibfnamefont {C.~G.}\ \bibnamefont
  {Durfee}}, \bibinfo {author} {\bibfnamefont {J.}~\bibnamefont {Lynch}}, \
  and\ \bibinfo {author} {\bibfnamefont {H.~M.}\ \bibnamefont {Milchberg}},\
  }\href {\doibase 10.1103/PhysRevE.51.2368} {\bibfield  {journal} {\bibinfo
  {journal} {Phys Rev E}\ }\textbf {\bibinfo {volume} {51}},\ \bibinfo {pages}
  {2368} (\bibinfo {year} {1995})}\BibitemShut {NoStop}%
\bibitem [{\citenamefont {Kumarappan}\ \emph {et~al.}(2005)\citenamefont
  {Kumarappan}, \citenamefont {Kim},\ and\ \citenamefont
  {Milchberg}}]{Kumarappan:2005du}%
  \BibitemOpen
  \bibfield  {author} {\bibinfo {author} {\bibfnamefont {V.}~\bibnamefont
  {Kumarappan}}, \bibinfo {author} {\bibfnamefont {K.}~\bibnamefont {Kim}}, \
  and\ \bibinfo {author} {\bibfnamefont {H.}~\bibnamefont {Milchberg}},\ }\href
  {\doibase 10.1103/PhysRevLett.94.205004} {\bibfield  {journal} {\bibinfo
  {journal} {Phys Rev Lett}\ }\textbf {\bibinfo {volume} {94}},\ \bibinfo
  {pages} {205004} (\bibinfo {year} {2005})}\BibitemShut {NoStop}%
\bibitem [{\citenamefont {Spence}\ and\ \citenamefont
  {Hooker}(2000)}]{Spence:2000fr}%
  \BibitemOpen
  \bibfield  {author} {\bibinfo {author} {\bibfnamefont {D.}~\bibnamefont
  {Spence}}\ and\ \bibinfo {author} {\bibfnamefont {S.}~\bibnamefont
  {Hooker}},\ }\href {\doibase 10.1103/PhysRevE.63.015401} {\bibfield
  {journal} {\bibinfo  {journal} {Phys Rev E}\ }\textbf {\bibinfo {volume}
  {63}} (\bibinfo {year} {2000}),\ 10.1103/PhysRevE.63.015401}\BibitemShut
  {NoStop}%
\bibitem [{\citenamefont {Butler}\ \emph {et~al.}(2002)\citenamefont {Butler},
  \citenamefont {Spence},\ and\ \citenamefont {Hooker}}]{Butler:2002zza}%
  \BibitemOpen
  \bibfield  {author} {\bibinfo {author} {\bibfnamefont {A.}~\bibnamefont
  {Butler}}, \bibinfo {author} {\bibfnamefont {D.~J.}\ \bibnamefont {Spence}},
  \ and\ \bibinfo {author} {\bibfnamefont {S.~M.}\ \bibnamefont {Hooker}},\
  }\href {\doibase 10.1103/PhysRevLett.89.185003} {\bibfield  {journal}
  {\bibinfo  {journal} {Phys Rev Lett}\ }\textbf {\bibinfo {volume} {89}},\
  \bibinfo {pages} {185003} (\bibinfo {year} {2002})}\BibitemShut {NoStop}%
\bibitem [{\citenamefont {Geddes}\ \emph {et~al.}(2004)\citenamefont {Geddes},
  \citenamefont {Toth}, \citenamefont {van Tilborg~J}, \citenamefont {Esarey},
  \citenamefont {Schroeder}, \citenamefont {Bruhwiler}, \citenamefont {Nieter},
  \citenamefont {Cary},\ and\ \citenamefont {Leemans}}]{Geddes:2004}%
  \BibitemOpen
  \bibfield  {author} {\bibinfo {author} {\bibfnamefont {C.~G.~R.}\
  \bibnamefont {Geddes}}, \bibinfo {author} {\bibfnamefont {C.}~\bibnamefont
  {Toth}}, \bibinfo {author} {\bibnamefont {van Tilborg~J}}, \bibinfo {author}
  {\bibfnamefont {E.}~\bibnamefont {Esarey}}, \bibinfo {author} {\bibfnamefont
  {C.~B.}\ \bibnamefont {Schroeder}}, \bibinfo {author} {\bibfnamefont
  {D.}~\bibnamefont {Bruhwiler}}, \bibinfo {author} {\bibfnamefont
  {C.}~\bibnamefont {Nieter}}, \bibinfo {author} {\bibfnamefont
  {J.}~\bibnamefont {Cary}}, \ and\ \bibinfo {author} {\bibfnamefont {W.~P.}\
  \bibnamefont {Leemans}},\ }\href {\doibase 10.1038/nature02900} {\bibfield
  {journal} {\bibinfo  {journal} {Nature}\ }\textbf {\bibinfo {volume} {431}},\
  \bibinfo {pages} {538} (\bibinfo {year} {2004})}\BibitemShut {NoStop}%
\bibitem [{\citenamefont {Karsch}\ \emph {et~al.}(2007)\citenamefont {Karsch},
  \citenamefont {Osterhoff}, \citenamefont {Popp}, \citenamefont
  {Rowlands-Rees}, \citenamefont {Major}, \citenamefont {Fuchs}, \citenamefont
  {Marx}, \citenamefont {Hoerlein}, \citenamefont {Schmid}, \citenamefont
  {Veisz}, \citenamefont {Becker}, \citenamefont {Schramm}, \citenamefont
  {Hidding}, \citenamefont {Pretzler}, \citenamefont {Habs}, \citenamefont
  {Gruener}, \citenamefont {Krausz},\ and\ \citenamefont
  {Hooker}}]{Karsch:2007}%
  \BibitemOpen
  \bibfield  {author} {\bibinfo {author} {\bibfnamefont {S.}~\bibnamefont
  {Karsch}}, \bibinfo {author} {\bibfnamefont {J.}~\bibnamefont {Osterhoff}},
  \bibinfo {author} {\bibfnamefont {A.}~\bibnamefont {Popp}}, \bibinfo {author}
  {\bibfnamefont {T.~P.}\ \bibnamefont {Rowlands-Rees}}, \bibinfo {author}
  {\bibfnamefont {Z.}~\bibnamefont {Major}}, \bibinfo {author} {\bibfnamefont
  {M.}~\bibnamefont {Fuchs}}, \bibinfo {author} {\bibfnamefont
  {B.}~\bibnamefont {Marx}}, \bibinfo {author} {\bibfnamefont {R.}~\bibnamefont
  {Hoerlein}}, \bibinfo {author} {\bibfnamefont {K.}~\bibnamefont {Schmid}},
  \bibinfo {author} {\bibfnamefont {L.}~\bibnamefont {Veisz}}, \bibinfo
  {author} {\bibfnamefont {S.}~\bibnamefont {Becker}}, \bibinfo {author}
  {\bibfnamefont {U.}~\bibnamefont {Schramm}}, \bibinfo {author} {\bibfnamefont
  {B.}~\bibnamefont {Hidding}}, \bibinfo {author} {\bibfnamefont
  {G.}~\bibnamefont {Pretzler}}, \bibinfo {author} {\bibfnamefont
  {D.}~\bibnamefont {Habs}}, \bibinfo {author} {\bibfnamefont {F.}~\bibnamefont
  {Gruener}}, \bibinfo {author} {\bibfnamefont {F.}~\bibnamefont {Krausz}}, \
  and\ \bibinfo {author} {\bibfnamefont {S.~M.}\ \bibnamefont {Hooker}},\
  }\href {\doibase 10.1088/1367-2630/9/11/415} {\bibfield  {journal} {\bibinfo
  {journal} {New J Phys}\ }\textbf {\bibinfo {volume} {9}},\ \bibinfo {pages}
  {415} (\bibinfo {year} {2007})}\BibitemShut {NoStop}%
\bibitem [{\citenamefont {Ibbotson}\ \emph {et~al.}(2010)\citenamefont
  {Ibbotson}, \citenamefont {Bourgeois}, \citenamefont {Rowlands-Rees},
  \citenamefont {Caballero}, \citenamefont {Bajlekov}, \citenamefont {Walker},
  \citenamefont {Kneip}, \citenamefont {Mangles}, \citenamefont {Nagel},
  \citenamefont {Palmer}, \citenamefont {Delerue}, \citenamefont {Doucas},
  \citenamefont {Urner}, \citenamefont {Chekhlov}, \citenamefont {Clarke},
  \citenamefont {Divall}, \citenamefont {Ertel}, \citenamefont {Foster},
  \citenamefont {Hawkes}, \citenamefont {Hooker}, \citenamefont {Parry},
  \citenamefont {Rajeev}, \citenamefont {Streeter},\ and\ \citenamefont
  {Hooker}}]{Ibbotson:2010}%
  \BibitemOpen
  \bibfield  {author} {\bibinfo {author} {\bibfnamefont {T.~P.~A.}\
  \bibnamefont {Ibbotson}}, \bibinfo {author} {\bibfnamefont {N.}~\bibnamefont
  {Bourgeois}}, \bibinfo {author} {\bibfnamefont {T.~P.}\ \bibnamefont
  {Rowlands-Rees}}, \bibinfo {author} {\bibfnamefont {L.~S.}\ \bibnamefont
  {Caballero}}, \bibinfo {author} {\bibfnamefont {S.~I.}\ \bibnamefont
  {Bajlekov}}, \bibinfo {author} {\bibfnamefont {P.~A.}\ \bibnamefont
  {Walker}}, \bibinfo {author} {\bibfnamefont {S.}~\bibnamefont {Kneip}},
  \bibinfo {author} {\bibfnamefont {S.~P.~D.}\ \bibnamefont {Mangles}},
  \bibinfo {author} {\bibfnamefont {S.~R.}\ \bibnamefont {Nagel}}, \bibinfo
  {author} {\bibfnamefont {C.~A.~J.}\ \bibnamefont {Palmer}}, \bibinfo {author}
  {\bibfnamefont {N.}~\bibnamefont {Delerue}}, \bibinfo {author} {\bibfnamefont
  {G.}~\bibnamefont {Doucas}}, \bibinfo {author} {\bibfnamefont
  {D.}~\bibnamefont {Urner}}, \bibinfo {author} {\bibfnamefont
  {O.}~\bibnamefont {Chekhlov}}, \bibinfo {author} {\bibfnamefont {R.~J.}\
  \bibnamefont {Clarke}}, \bibinfo {author} {\bibfnamefont {E.}~\bibnamefont
  {Divall}}, \bibinfo {author} {\bibfnamefont {K.}~\bibnamefont {Ertel}},
  \bibinfo {author} {\bibfnamefont {P.~S.}\ \bibnamefont {Foster}}, \bibinfo
  {author} {\bibfnamefont {S.~J.}\ \bibnamefont {Hawkes}}, \bibinfo {author}
  {\bibfnamefont {C.~J.}\ \bibnamefont {Hooker}}, \bibinfo {author}
  {\bibfnamefont {B.}~\bibnamefont {Parry}}, \bibinfo {author} {\bibfnamefont
  {P.~P.}\ \bibnamefont {Rajeev}}, \bibinfo {author} {\bibfnamefont {M.~J.~V.}\
  \bibnamefont {Streeter}}, \ and\ \bibinfo {author} {\bibfnamefont {S.~M.}\
  \bibnamefont {Hooker}},\ }\href {\doibase 10.1103/PhysRevSTAB.13.031301}
  {\bibfield  {journal} {\bibinfo  {journal} {Phys. Rev. ST Accel. Beams}\
  }\textbf {\bibinfo {volume} {13}},\ \bibinfo {pages} {031301} (\bibinfo
  {year} {2010})}\BibitemShut {NoStop}%
\bibitem [{\citenamefont {Rowlands-Rees}\ \emph {et~al.}(2008)\citenamefont
  {Rowlands-Rees}, \citenamefont {Kamperidis}, \citenamefont {Kneip},
  \citenamefont {Gonsalves}, \citenamefont {Mangles}, \citenamefont
  {Gallacher}, \citenamefont {Brunetti}, \citenamefont {Ibbotson},
  \citenamefont {Murphy}, \citenamefont {Foster}, \citenamefont {Streeter},
  \citenamefont {Budde}, \citenamefont {Norreys}, \citenamefont {Jaroszynski},
  \citenamefont {Krushelnick}, \citenamefont {Najmudin},\ and\ \citenamefont
  {Hooker}}]{Rowlands-Rees:2008}%
  \BibitemOpen
  \bibfield  {author} {\bibinfo {author} {\bibfnamefont {T.~P.}\ \bibnamefont
  {Rowlands-Rees}}, \bibinfo {author} {\bibfnamefont {C.}~\bibnamefont
  {Kamperidis}}, \bibinfo {author} {\bibfnamefont {S.}~\bibnamefont {Kneip}},
  \bibinfo {author} {\bibfnamefont {A.~J.}\ \bibnamefont {Gonsalves}}, \bibinfo
  {author} {\bibfnamefont {S.~P.~D.}\ \bibnamefont {Mangles}}, \bibinfo
  {author} {\bibfnamefont {J.~G.}\ \bibnamefont {Gallacher}}, \bibinfo {author}
  {\bibfnamefont {E.}~\bibnamefont {Brunetti}}, \bibinfo {author}
  {\bibfnamefont {T.}~\bibnamefont {Ibbotson}}, \bibinfo {author}
  {\bibfnamefont {C.~D.}\ \bibnamefont {Murphy}}, \bibinfo {author}
  {\bibfnamefont {P.~S.}\ \bibnamefont {Foster}}, \bibinfo {author}
  {\bibfnamefont {M.~J.~V.}\ \bibnamefont {Streeter}}, \bibinfo {author}
  {\bibfnamefont {F.}~\bibnamefont {Budde}}, \bibinfo {author} {\bibfnamefont
  {P.~A.}\ \bibnamefont {Norreys}}, \bibinfo {author} {\bibfnamefont {D.~A.}\
  \bibnamefont {Jaroszynski}}, \bibinfo {author} {\bibfnamefont
  {K.}~\bibnamefont {Krushelnick}}, \bibinfo {author} {\bibfnamefont
  {Z.}~\bibnamefont {Najmudin}}, \ and\ \bibinfo {author} {\bibfnamefont
  {S.~M.}\ \bibnamefont {Hooker}},\ }\href {\doibase
  10.1103/PhysRevLett.100.105005} {\bibfield  {journal} {\bibinfo  {journal}
  {Phys Rev Lett}\ }\textbf {\bibinfo {volume} {100}},\ \bibinfo {pages}
  {105005} (\bibinfo {year} {2008})}\BibitemShut {NoStop}%
\bibitem [{\citenamefont {Goers}\ \emph {et~al.}(2014)\citenamefont {Goers},
  \citenamefont {Yoon}, \citenamefont {Elle}, \citenamefont {Hine},\ and\
  \citenamefont {Milchberg}}]{Goers:2014jg}%
  \BibitemOpen
  \bibfield  {author} {\bibinfo {author} {\bibfnamefont {A.~J.}\ \bibnamefont
  {Goers}}, \bibinfo {author} {\bibfnamefont {S.~J.}\ \bibnamefont {Yoon}},
  \bibinfo {author} {\bibfnamefont {J.~A.}\ \bibnamefont {Elle}}, \bibinfo
  {author} {\bibfnamefont {G.~A.}\ \bibnamefont {Hine}}, \ and\ \bibinfo
  {author} {\bibfnamefont {H.~M.}\ \bibnamefont {Milchberg}},\ }\href {\doibase
  10.1063/1.4880102} {\bibfield  {journal} {\bibinfo  {journal} {Appl. Phys.
  Lett.}\ }\textbf {\bibinfo {volume} {104}},\ \bibinfo {pages} {214105}
  (\bibinfo {year} {2014})}\BibitemShut {NoStop}%
\bibitem [{\citenamefont {Gonsalves}\ \emph {et~al.}(2016)\citenamefont
  {Gonsalves}, \citenamefont {Liu}, \citenamefont {Bobrova}, \citenamefont
  {Sasorov}, \citenamefont {Pieronek}, \citenamefont {Daniels}, \citenamefont
  {Antipov}, \citenamefont {Butler}, \citenamefont {Bulanov}, \citenamefont
  {Waldron}, \citenamefont {Mittelberger},\ and\ \citenamefont
  {Leemans}}]{Gonsalves:2016jc}%
  \BibitemOpen
  \bibfield  {author} {\bibinfo {author} {\bibfnamefont {A.~J.}\ \bibnamefont
  {Gonsalves}}, \bibinfo {author} {\bibfnamefont {F.}~\bibnamefont {Liu}},
  \bibinfo {author} {\bibfnamefont {N.~A.}\ \bibnamefont {Bobrova}}, \bibinfo
  {author} {\bibfnamefont {P.~V.}\ \bibnamefont {Sasorov}}, \bibinfo {author}
  {\bibfnamefont {C.}~\bibnamefont {Pieronek}}, \bibinfo {author}
  {\bibfnamefont {J.}~\bibnamefont {Daniels}}, \bibinfo {author} {\bibfnamefont
  {S.}~\bibnamefont {Antipov}}, \bibinfo {author} {\bibfnamefont {J.~E.}\
  \bibnamefont {Butler}}, \bibinfo {author} {\bibfnamefont {S.~S.}\
  \bibnamefont {Bulanov}}, \bibinfo {author} {\bibfnamefont {W.~L.}\
  \bibnamefont {Waldron}}, \bibinfo {author} {\bibfnamefont {D.~E.}\
  \bibnamefont {Mittelberger}}, \ and\ \bibinfo {author} {\bibfnamefont
  {W.~P.}\ \bibnamefont {Leemans}},\ }\href {\doibase 10.1063/1.4940121}
  {\bibfield  {journal} {\bibinfo  {journal} {J Appl Phys}\ }\textbf {\bibinfo
  {volume} {119}},\ \bibinfo {pages} {033302} (\bibinfo {year}
  {2016})}\BibitemShut {NoStop}%
\bibitem [{\citenamefont {Clark}\ and\ \citenamefont
  {Milchberg}(1997)}]{Clark:1997we}%
  \BibitemOpen
  \bibfield  {author} {\bibinfo {author} {\bibfnamefont {T.~R.}\ \bibnamefont
  {Clark}}\ and\ \bibinfo {author} {\bibfnamefont {H.~M.}\ \bibnamefont
  {Milchberg}},\ }\href
  {http://journals.aps.org/prl/abstract/10.1103/PhysRevLett.78.2373} {\bibfield
   {journal} {\bibinfo  {journal} {Phys Rev Lett}\ }\textbf {\bibinfo {volume}
  {78}},\ \bibinfo {pages} {2373} (\bibinfo {year} {1997})}\BibitemShut
  {NoStop}%
\bibitem [{\citenamefont {Clark}\ and\ \citenamefont
  {Milchberg}(2000)}]{Clark:2000dk}%
  \BibitemOpen
  \bibfield  {author} {\bibinfo {author} {\bibfnamefont {T.~R.}\ \bibnamefont
  {Clark}}\ and\ \bibinfo {author} {\bibfnamefont {H.~M.}\ \bibnamefont
  {Milchberg}},\ }\href {\doibase 10.1103/PhysRevE.61.1954} {\bibfield
  {journal} {\bibinfo  {journal} {Phys Rev E}\ }\textbf {\bibinfo {volume}
  {61}},\ \bibinfo {pages} {1954} (\bibinfo {year} {2000})}\BibitemShut
  {NoStop}%
\bibitem [{\citenamefont {Lemos}\ \emph
  {et~al.}(2013{\natexlab{a}})\citenamefont {Lemos}, \citenamefont {Grismayer},
  \citenamefont {Cardoso}, \citenamefont {Figueira}, \citenamefont {Issac},
  \citenamefont {Jaroszynski},\ and\ \citenamefont {Dias}}]{Lemos:2013gb}%
  \BibitemOpen
  \bibfield  {author} {\bibinfo {author} {\bibfnamefont {N.}~\bibnamefont
  {Lemos}}, \bibinfo {author} {\bibfnamefont {T.}~\bibnamefont {Grismayer}},
  \bibinfo {author} {\bibfnamefont {L.}~\bibnamefont {Cardoso}}, \bibinfo
  {author} {\bibfnamefont {G.}~\bibnamefont {Figueira}}, \bibinfo {author}
  {\bibfnamefont {R.}~\bibnamefont {Issac}}, \bibinfo {author} {\bibfnamefont
  {D.~A.}\ \bibnamefont {Jaroszynski}}, \ and\ \bibinfo {author} {\bibfnamefont
  {J.~M.}\ \bibnamefont {Dias}},\ }\href {\doibase 10.1063/1.4810797}
  {\bibfield  {journal} {\bibinfo  {journal} {Phys Plasmas}\ }\textbf {\bibinfo
  {volume} {20}},\ \bibinfo {pages} {063102} (\bibinfo {year}
  {2013}{\natexlab{a}})}\BibitemShut {NoStop}%
\bibitem [{\citenamefont {Lemos}\ \emph
  {et~al.}(2013{\natexlab{b}})\citenamefont {Lemos}, \citenamefont {Grismayer},
  \citenamefont {Cardoso}, \citenamefont {Geada}, \citenamefont {Figueira},\
  and\ \citenamefont {Dias}}]{Lemos:2013ju}%
  \BibitemOpen
  \bibfield  {author} {\bibinfo {author} {\bibfnamefont {N.}~\bibnamefont
  {Lemos}}, \bibinfo {author} {\bibfnamefont {T.}~\bibnamefont {Grismayer}},
  \bibinfo {author} {\bibfnamefont {L.}~\bibnamefont {Cardoso}}, \bibinfo
  {author} {\bibfnamefont {J.}~\bibnamefont {Geada}}, \bibinfo {author}
  {\bibfnamefont {G.}~\bibnamefont {Figueira}}, \ and\ \bibinfo {author}
  {\bibfnamefont {J.~M.}\ \bibnamefont {Dias}},\ }\href {\doibase
  10.1063/1.4825228} {\bibfield  {journal} {\bibinfo  {journal} {Phys Plasmas}\
  }\textbf {\bibinfo {volume} {20}},\ \bibinfo {pages} {103109} (\bibinfo
  {year} {2013}{\natexlab{b}})}\BibitemShut {NoStop}%
\bibitem [{\citenamefont {Corkum}\ \emph {et~al.}(1989)\citenamefont {Corkum},
  \citenamefont {Burnett},\ and\ \citenamefont {Brunel}}]{Corkum:1989ji}%
  \BibitemOpen
  \bibfield  {author} {\bibinfo {author} {\bibfnamefont {P.~B.}\ \bibnamefont
  {Corkum}}, \bibinfo {author} {\bibfnamefont {N.~H.}\ \bibnamefont {Burnett}},
  \ and\ \bibinfo {author} {\bibfnamefont {F.}~\bibnamefont {Brunel}},\ }\href
  {\doibase 10.1103/PhysRevLett.62.1259} {\bibfield  {journal} {\bibinfo
  {journal} {Phys Rev Lett}\ }\textbf {\bibinfo {volume} {62}},\ \bibinfo
  {pages} {1259} (\bibinfo {year} {1989})}\BibitemShut {NoStop}%
\bibitem [{\citenamefont {Gallagher}(1988)}]{Gallagher:1988er}%
  \BibitemOpen
  \bibfield  {author} {\bibinfo {author} {\bibfnamefont {T.~F.}\ \bibnamefont
  {Gallagher}},\ }\href {\doibase 10.1103/PhysRevLett.61.2304} {\bibfield
  {journal} {\bibinfo  {journal} {Phys Rev Lett}\ }\textbf {\bibinfo {volume}
  {61}},\ \bibinfo {pages} {2304} (\bibinfo {year} {1988})}\BibitemShut
  {NoStop}%
\bibitem [{\citenamefont {Arber}\ \emph {et~al.}(2015)\citenamefont {Arber},
  \citenamefont {Bennett}, \citenamefont {Brady}, \citenamefont
  {Lawrence-Douglas}, \citenamefont {Ramsay}, \citenamefont {Sircombe},
  \citenamefont {Gillies}, \citenamefont {Evans}, \citenamefont {Schmitz},
  \citenamefont {Bell},\ and\ \citenamefont {Ridgers}}]{Arber2015}%
  \BibitemOpen
  \bibfield  {author} {\bibinfo {author} {\bibfnamefont {T.~D.}\ \bibnamefont
  {Arber}}, \bibinfo {author} {\bibfnamefont {K.}~\bibnamefont {Bennett}},
  \bibinfo {author} {\bibfnamefont {C.~S.}\ \bibnamefont {Brady}}, \bibinfo
  {author} {\bibfnamefont {A.}~\bibnamefont {Lawrence-Douglas}}, \bibinfo
  {author} {\bibfnamefont {M.~G.}\ \bibnamefont {Ramsay}}, \bibinfo {author}
  {\bibfnamefont {N.~J.}\ \bibnamefont {Sircombe}}, \bibinfo {author}
  {\bibfnamefont {P.}~\bibnamefont {Gillies}}, \bibinfo {author} {\bibfnamefont
  {R.~G.}\ \bibnamefont {Evans}}, \bibinfo {author} {\bibfnamefont
  {H.}~\bibnamefont {Schmitz}}, \bibinfo {author} {\bibfnamefont {A.~R.}\
  \bibnamefont {Bell}}, \ and\ \bibinfo {author} {\bibfnamefont {C.~P.}\
  \bibnamefont {Ridgers}},\ }\href {\doibase 10.1088/0741-3335/57/11/113001}
  {\bibfield  {journal} {\bibinfo  {journal} {Plasma Phys. Control. Fusion}\
  }\textbf {\bibinfo {volume} {57}},\ \bibinfo {pages} {113001} (\bibinfo
  {year} {2015})}\BibitemShut {NoStop}%
\bibitem [{\citenamefont {Posthumus}\ \emph {et~al.}(1997)\citenamefont
  {Posthumus}, \citenamefont {Thompson}, \citenamefont {Frasinki},\ and\
  \citenamefont {Codling}}]{Posthumus1997}%
  \BibitemOpen
  \bibfield  {author} {\bibinfo {author} {\bibfnamefont {J.}~\bibnamefont
  {Posthumus}}, \bibinfo {author} {\bibfnamefont {M.}~\bibnamefont {Thompson}},
  \bibinfo {author} {\bibfnamefont {I.}~\bibnamefont {Frasinki}}, \ and\
  \bibinfo {author} {\bibfnamefont {K.}~\bibnamefont {Codling}},\ }in\
  \href@noop {} {\emph {\bibinfo {booktitle} {Inst. Phys. Conf. Ser. No.
  154}}}\ (\bibinfo  {publisher} {IOP Publishing Ltd},\ \bibinfo {year}
  {1997})\ p.\ \bibinfo {pages} {298}\BibitemShut {NoStop}%
\bibitem [{\citenamefont {Lawrence-Douglas}(2013)}]{Lawrence-Douglas2013}%
  \BibitemOpen
  \bibfield  {author} {\bibinfo {author} {\bibfnamefont {A.}~\bibnamefont
  {Lawrence-Douglas}},\ }\emph {\bibinfo {title} {{Ionisation Effects for
  Laser-Plasma Interactions by Particle-in-Cell Code}}},\ \href
  {http://webcat.warwick.ac.uk/record=b2686775~S1} {Ph.D. thesis},\ \bibinfo
  {school} {University of Warwick} (\bibinfo {year} {2013})\BibitemShut
  {NoStop}%
\bibitem [{\citenamefont {Tauscher}\ \emph {et~al.}(2017)\citenamefont
  {Tauscher}, \citenamefont {Schaper}, \citenamefont {Bohlen}, \citenamefont
  {Poder}, \citenamefont {Schwinkendorf}, \citenamefont {Mehrling},
  \citenamefont {Wesch}, \citenamefont {Goldberg}, \citenamefont {Quast},
  \citenamefont {Aschikhin}, \citenamefont {Roeckemann}, \citenamefont {Dale},
  \citenamefont {Streeter},\ and\ \citenamefont {Osterhoff}}]{Tauscher:2017}%
  \BibitemOpen
  \bibfield  {author} {\bibinfo {author} {\bibfnamefont {G.}~\bibnamefont
  {Tauscher}}, \bibinfo {author} {\bibfnamefont {L.}~\bibnamefont {Schaper}},
  \bibinfo {author} {\bibfnamefont {S.}~\bibnamefont {Bohlen}}, \bibinfo
  {author} {\bibfnamefont {K.}~\bibnamefont {Poder}}, \bibinfo {author}
  {\bibfnamefont {J.-P.}\ \bibnamefont {Schwinkendorf}}, \bibinfo {author}
  {\bibfnamefont {T.}~\bibnamefont {Mehrling}}, \bibinfo {author}
  {\bibfnamefont {S.}~\bibnamefont {Wesch}}, \bibinfo {author} {\bibfnamefont
  {L.}~\bibnamefont {Goldberg}}, \bibinfo {author} {\bibfnamefont
  {M.}~\bibnamefont {Quast}}, \bibinfo {author} {\bibfnamefont
  {A.}~\bibnamefont {Aschikhin}}, \bibinfo {author} {\bibfnamefont {J.-H.}\
  \bibnamefont {Roeckemann}}, \bibinfo {author} {\bibfnamefont
  {J.}~\bibnamefont {Dale}}, \bibinfo {author} {\bibfnamefont {M.}~\bibnamefont
  {Streeter}}, \ and\ \bibinfo {author} {\bibfnamefont {J.}~\bibnamefont
  {Osterhoff}}\ }(\bibinfo {organization} {3rd European Advanced Accelerator
  Concepts Workshop, La Biodola, Isola d'Elba (Italy), 24 Sep 2017 - 30 Sep
  2017},\ \bibinfo {year} {2017})\BibitemShut {NoStop}%
\bibitem [{\citenamefont {{Shiner}}\ \emph {et~al.}(1993)\citenamefont
  {{Shiner}}, \citenamefont {{Gilligan}}, \citenamefont {{Cook}},\ and\
  \citenamefont {{Lichten}}}]{Shiner:1993}%
  \BibitemOpen
  \bibfield  {author} {\bibinfo {author} {\bibfnamefont {D.}~\bibnamefont
  {{Shiner}}}, \bibinfo {author} {\bibfnamefont {J.~M.}\ \bibnamefont
  {{Gilligan}}}, \bibinfo {author} {\bibfnamefont {B.~M.}\ \bibnamefont
  {{Cook}}}, \ and\ \bibinfo {author} {\bibfnamefont {W.}~\bibnamefont
  {{Lichten}}},\ }\href {\doibase 10.1103/PhysRevA.47.4042} {\bibfield
  {journal} {\bibinfo  {journal} {\pra}\ }\textbf {\bibinfo {volume} {47}},\
  \bibinfo {pages} {4042} (\bibinfo {year} {1993})}\BibitemShut {NoStop}%
\bibitem [{\citenamefont {Spitzer}(1967)}]{Spitzer1967}%
  \BibitemOpen
  \bibfield  {author} {\bibinfo {author} {\bibfnamefont {L.}~\bibnamefont
  {Spitzer}},\ }\href@noop {} {\emph {\bibinfo {title} {{Physics of Fully
  Ionized Gases}}}}\ (\bibinfo  {publisher} {Interscience Publishers},\
  \bibinfo {year} {1967})\BibitemShut {NoStop}%
\bibitem [{\citenamefont {MacFarlane}\ \emph {et~al.}(2006)\citenamefont
  {MacFarlane}, \citenamefont {Golovkin},\ and\ \citenamefont
  {Woodruff}}]{MacFarlane2006}%
  \BibitemOpen
  \bibfield  {author} {\bibinfo {author} {\bibfnamefont {J.~J.}\ \bibnamefont
  {MacFarlane}}, \bibinfo {author} {\bibfnamefont {I.~E.}\ \bibnamefont
  {Golovkin}}, \ and\ \bibinfo {author} {\bibfnamefont {P.~R.}\ \bibnamefont
  {Woodruff}},\ }\href {\doibase 10.1016/j.jqsrt.2005.05.031} {\bibfield
  {journal} {\bibinfo  {journal} {J. Quant. Spectrosc. Radiat. Transf.}\
  }\textbf {\bibinfo {volume} {99}},\ \bibinfo {pages} {381} (\bibinfo {year}
  {2006})}\BibitemShut {NoStop}%
\bibitem [{\citenamefont {Fan}\ \emph {et~al.}(2000)\citenamefont {Fan},
  \citenamefont {Parra},\ and\ \citenamefont {Milchberg}}]{Fan:2000fh}%
  \BibitemOpen
  \bibfield  {author} {\bibinfo {author} {\bibfnamefont {J.}~\bibnamefont
  {Fan}}, \bibinfo {author} {\bibfnamefont {E.}~\bibnamefont {Parra}}, \ and\
  \bibinfo {author} {\bibfnamefont {H.~M.}\ \bibnamefont {Milchberg}},\ }\href
  {\doibase 10.1103/PhysRevLett.84.3085} {\bibfield  {journal} {\bibinfo
  {journal} {Phys Rev Lett}\ }\textbf {\bibinfo {volume} {84}},\ \bibinfo
  {pages} {3085} (\bibinfo {year} {2000})}\BibitemShut {NoStop}%
\bibitem [{\citenamefont {Taylor}(1950)}]{Taylor1950}%
  \BibitemOpen
  \bibfield  {author} {\bibinfo {author} {\bibfnamefont {G.}~\bibnamefont
  {Taylor}},\ }\href@noop {} {\bibfield  {journal} {\bibinfo  {journal} {Proc.
  R. Soc. A}\ }\textbf {\bibinfo {volume} {201}},\ \bibinfo {pages} {159}
  (\bibinfo {year} {1950})}\BibitemShut {NoStop}%
\bibitem [{\citenamefont {Hutchens}(2000)}]{Hutchens1995}%
  \BibitemOpen
  \bibfield  {author} {\bibinfo {author} {\bibfnamefont {G.~J.}\ \bibnamefont
  {Hutchens}},\ }\href@noop {} {\bibfield  {journal} {\bibinfo  {journal}
  {Journal of Applied Physics}\ }\textbf {\bibinfo {volume} {88}},\ \bibinfo
  {pages} {3654} (\bibinfo {year} {2000})}\BibitemShut {NoStop}%
\bibitem [{\citenamefont {Takeda}\ \emph {et~al.}(1982)\citenamefont {Takeda},
  \citenamefont {Ina},\ and\ \citenamefont {Kobayashi}}]{Takeda:1982}%
  \BibitemOpen
  \bibfield  {author} {\bibinfo {author} {\bibfnamefont {M.}~\bibnamefont
  {Takeda}}, \bibinfo {author} {\bibfnamefont {H.}~\bibnamefont {Ina}}, \ and\
  \bibinfo {author} {\bibfnamefont {S.}~\bibnamefont {Kobayashi}},\ }\href
  {\doibase 10.1364/JOSA.72.000156} {\bibfield  {journal} {\bibinfo  {journal}
  {J. Opt. Soc. Am.}\ }\textbf {\bibinfo {volume} {72}},\ \bibinfo {pages}
  {156} (\bibinfo {year} {1982})}\BibitemShut {NoStop}%
\bibitem [{\citenamefont {Bone}\ \emph {et~al.}(1986)\citenamefont {Bone},
  \citenamefont {Bachor},\ and\ \citenamefont {Sandeman}}]{Bone:1986}%
  \BibitemOpen
  \bibfield  {author} {\bibinfo {author} {\bibfnamefont {D.~J.}\ \bibnamefont
  {Bone}}, \bibinfo {author} {\bibfnamefont {H.-A.}\ \bibnamefont {Bachor}}, \
  and\ \bibinfo {author} {\bibfnamefont {R.~J.}\ \bibnamefont {Sandeman}},\
  }\href {\doibase 10.1364/AO.25.001653} {\bibfield  {journal} {\bibinfo
  {journal} {Appl. Opt.}\ }\textbf {\bibinfo {volume} {25}},\ \bibinfo {pages}
  {1653} (\bibinfo {year} {1986})}\BibitemShut {NoStop}%
\bibitem [{\citenamefont {McLeod}(1954)}]{McLeod:1954}%
  \BibitemOpen
  \bibfield  {author} {\bibinfo {author} {\bibfnamefont {J.~H.}\ \bibnamefont
  {McLeod}},\ }\href {\doibase 10.1364/JOSA.44.000592} {\bibfield  {journal}
  {\bibinfo  {journal} {J. Opt. Soc. Am.}\ }\textbf {\bibinfo {volume} {44}},\
  \bibinfo {pages} {592} (\bibinfo {year} {1954})}\BibitemShut {NoStop}%
\bibitem [{\citenamefont {Sochacki}\ \emph {et~al.}(1992)\citenamefont
  {Sochacki}, \citenamefont {Bar{\'{a}}}, \citenamefont {Jaroszewicz},\ and\
  \citenamefont {Ko{\l}odziejczyk}}]{Sochacki:1992}%
  \BibitemOpen
  \bibfield  {author} {\bibinfo {author} {\bibfnamefont {J.}~\bibnamefont
  {Sochacki}}, \bibinfo {author} {\bibfnamefont {S.}~\bibnamefont
  {Bar{\'{a}}}}, \bibinfo {author} {\bibfnamefont {Z.}~\bibnamefont
  {Jaroszewicz}}, \ and\ \bibinfo {author} {\bibfnamefont {A.}~\bibnamefont
  {Ko{\l}odziejczyk}},\ }\href {\doibase 10.1364/OL.17.000007} {\bibfield
  {journal} {\bibinfo  {journal} {Opt. Lett.}\ }\textbf {\bibinfo {volume}
  {17}},\ \bibinfo {pages} {7} (\bibinfo {year} {1992})}\BibitemShut {NoStop}%
\end{thebibliography}%

\end{document}